\documentclass[a4paper,10pt]{article}
\usepackage[utf8]{inputenc}
\usepackage{amsfonts}
\usepackage{amsthm}
\usepackage{amsmath}
\usepackage{amssymb}
\usepackage{setspace}
\usepackage{graphicx}
\usepackage{authblk}
\usepackage{array}
\usepackage{setspace}
\usepackage[margin=2.5cm]{geometry}
\numberwithin{equation}{section}
\usepackage{longtable}
\usepackage{xcolor}

\theoremstyle{definition}

\usepackage{appendix}
\theoremstyle{remark}

\numberwithin{equation}{section}
\newcommand{\ndv}[1]{\hat{#1}}



\title{Deducing Flux from Single Point Temperature History when Relative Spatial Variation of Flux is Prescribed}
\author[1]{David Buttsworth}
\author[2]{Timothy Buttsworth}
\affil[1]{School of Mechanical and Electrical Engineering, University of Southern Queensland}
\affil[2]{Department of Mathematics, Cornell University}

\date{April 30, 2020}

\begin{document}

\maketitle

\begin{abstract}
Surface heat transfer in convective and radiative environments is sometimes measured by recording the surface temperature history in a transient experiment and interpreting this surface temperature with the aid of a suitable model for transient conduction within the substrate.  The semi-infinite one-dimensional model is often adopted, and several well-developed techniques for application of this model to surface temperature data are available.  However, when a spatial variation of heat flux exists across the surface, the application of the semi-infinite one-dimensional approach may not always be a reasonable approximation.  In this paper we introduce a method for treatment of the measured surface temperature history that is more accurate than the semi-infinite one-dimensional approximation when substrate lateral conduction is significant and the relative spatial distribution of the flux is known \textit{a priori}.  This new method uses the so-called \textit{Neumann heat kernel}, which evolves a temperature over an insulated domain with unit energy initially deposited at a specified point. A useful impulse response function is formed by integrating this Neumann heat kernel against the spatial variation of flux over the surface of the domain. Neumann heat kernels are constructed for the solid box, cylinder, and sphere.  By applying the heat kernel result for the sphere to the analysis of a convective experiment using hemispherical-nosed probes, we demonstrate how the theoretical results enhance the practical analysis of transient surface temperature measurements.  The current approach is superior to former methods relying on semi-empirical approximations because the multi-dimensional heat conduction within the substrate is modelled with greater fidelity using the heat kernel analysis. 
\end{abstract}

\section{Introduction}

Measurement of heat flux in short duration wind tunnel experiments is often achieved using a transient approach in which the initial surface temperature of the test object is different from the flow temperature.  The change of the temperature of the object surface is then recorded as a function of time from the flow onset, and the heat flux is deduced using a theoretical model for heat conduction within the object substrate. The early development of transient heat transfer measurement techniques is discussed by Schultz and Jones \cite{Schultz1973}.  

If the heat flux is uniform across the object surface, then for some relatively short period of time, the heat flux can be deduced directly from the surface temperature measurement on the assumption of one-dimensional heat flow in the object substrate. However, in the more general case where the non-uniformity of the heat flux across the object surface is significant, either the surface temperature must be measured at multiple locations and some form of iterative or inverse multi-dimensional analysis must be used to deduce the heat flux, or some form of correction must be applied to the one-dimensional heat flow analysis to accommodate the lateral conduction.  While such approaches can be effective, they can be difficult to implement and they can also introduce noise and/or uncertainties through imprecise implementation or inherent approximations. 

The problem considered in the present work is the special case of a non-uniform heat flux where the relative spatial distribution of heat flux is prescribed.  Although the relative spatial distribution of heat flux is not always known prior to a given experiment, this special case still has practical relevance. For example, in the context of convective heat transfer experiments, the relative spatial variation of heat flux can be confidently prescribed under certain conditions when the flow is quasi-steady and the boundary layer is laminar.  To avoid the difficulties associated with iterative or inverse approaches in such cases, we show how the heat flux can be determined directly from measurement of surface temperature history at a single point. 

\section{Review of Related Work}

\subsection{One-Dimensional Analysis}

Temperature measurements may be taken at a modest number of locations across the test object via discrete gauges based on resistance thermometers, thermocouples, and fibre optic sensors.  Alternatively, temperature may be measured across broader areas of the test object surface through optical techniques such as infrared thermography and the application of special coating such as Thermochromic Liquid Crystal (TLC) or temperature-sensitive paint.  In all cases however, the heat transfer between the flow and the object surface or heat flux gauge is deduced from the recorded surface temperature history using a theoretical analysis for the heat conduction within the substrate. 

\subsubsection{Flat Surfaces}

The simplest models for the transient heat conduction treat the surface of the substrate as being flat.  If the temperature of a relatively thin layer of material on the test object or heat flux gauge surface is measured, then the heat flux may be deduced using a \emph{calorimetric} transient analysis. Alternatively, if the test object has sufficient thickness such that the majority of the temperature variation is confined to a relatively small zone near the surface of the object, then the \emph{semi-infinite} transient analysis may be an appropriate method for deducing the heat flux from the measured surface temperature.

For calorimetric measurements of a spatially uniform surface heat flux under the assumption of constant thermal properties of the test object or heat flux gauge -- constant values of density $\rho$ and specific heat $c_p$ -- the instantaneous heat flux from the calorimeter $q_{cal}(t)$ can be determined from the temperature history $T(t)$ of the thin layer of constant thickness $l$ using
\begin{equation}\label{eq:Cal_1D_heatConduction}
q_{cal} = \rho c_p l \frac{dT}{dt} ,
\end{equation}
provided there are no conduction losses or gains by the layer through supporting structures or backing materials.  Additional challenges in implementing this technique include the requirement for the temperature sensor to provide a true measurement of the mean temperature of the thin layer, and the temporal differentiation of the temperature measurement which will have a non-zero noise level.   

In the case of the semi-infinite approach, if the heat flux is uniform across the surface of test object, then the one-dimensional heat conduction model can be applied.  When the substrate surface is flat and the thermal properties of the substrate -- the density $\rho$, specific heat $c$, and conductivity $k$ -- are constant, the instantaneous heat transfer from the semi-infinite analysis $q_{si,fs}(t)$ can be determined from the measured temperature $T(t)$ using
\begin{equation}\label{eq:SI_1D_heatConduction}
q_{si,fs} = \frac{\sqrt{\rho c k}}{\sqrt{\pi}} \int_0^t \frac{dT}{d\tau} \frac{1}{\sqrt{t-\tau}} d\tau .
\end{equation}
A variety of schemes can be used to approximate $q_{si,fs}$ from (\ref{eq:SI_1D_heatConduction}) for the deduction of $q(t)$ from the measured temperature history $T(t)$ at discrete times.  While the Cook-Felderman \cite{Cook1966} approach has been used extensively, the linear time-invariant system digital filtering approach introduced by Oldfield \cite{Oldfield2008} offers speed and convenience and can be applied to multi-layered substrate cases as well, under the restriction of one-dimensional heat flow.
The Oldfield method is adapted to the multi-dimensional analysis performed in the present work.

\subsubsection{Curved Surfaces}

If the heat transfer surface is curved, then for relatively short periods of time the substrate may be treated as semi-infinite, but errors due to this approximation will increase with time.  An approximate analytical method to correct the semi-infinite analysis (\ref{eq:SI_1D_heatConduction}) for curvature of the surface was introduced in \cite{Buttsworth1997}, and the correction term for radial effects is given by
\begin{equation}\label{eq:q_r}
    q_r = -\frac{k \sigma}{2 R} \left( T - T_i \right)
\end{equation}
where $k$ is the thermal conductivity of the substrate, $\sigma = \pm 1$ for a cylinder and $\sigma = \pm 2$ for a sphere (the positive applying for a convex substrate, and the negative for a concave substrate), $R$ is the radius of the surface, and $T_i$ is the initial temperature of the surface.  The heat flux to the surface can then be calculated from a combination of the semi-infinite analysis and the radial correction as given by
\begin{equation}
    q = q_{si,fs} + q_r 
\end{equation}
For example, consider the case of a hot fluid flow that is suddenly initiated adjacent to a substrate that is approximated as concave cylinder.  For this case, $\sigma = -1$ and the heat flux to the surface that is inferred from a semi-infinite, flat surface analysis of the measured surface temperature according to (\ref{eq:SI_1D_heatConduction}) will underestimate the actual heat flux by the amount $q_r = \frac{k}{2R} (T-T_i)$.  In this concave geometry case, the relieving effect associated with the increase in the cross section area with increasing distance from the surface causes the surface temperature to rise at a lower rate relative to a flat surface that experiences the same heat flux. Therefore the quantity $q_r$ must be added to $q_{si,fs}$ to obtain a better estimate of the true heat flux at the surface.

A similar approach can also be adopted for analysis of experimental data where the heat transfer coefficient remains constant during the experiment as demonstrated in \cite{Buttsworth1997}, and Wagner et al. \cite{Wagner2005} performed further experiments demonstrating the validity and utility of the approximate analysis in the context of transient thermochromic liquid crystal (TLC) experiments. Wagner et al. \cite{Wagner2005} also presented an exact analysis which correctly accounts for finite substrate thickness effects in the case of an annulus subjected to convective boundary conditions on both surfaces.  In cases where finite substrate thickness effects are not significant, Wagner et al. \cite{Wagner2005} found the approximate analysis in \cite{Buttsworth1997} to be sufficiently accurate and less time consuming than the exact analysis in deducing the heat transfer coefficient from the TLC data. For the treatment of curvature effects in TLC experiments, Zhou et al. \cite{Zhou2019} sought a more accurate analysis for the heat transfer coefficient distributions through the solution of the inverse problem and such an approach introduces additional complexity to the method.

\subsection{Multi-Dimensional Analysis}

\subsubsection{Magnitude of Errors}

Although the assumption of one-dimensional heat flow within the test object or heat flux gauge (whether on a surface that is flat or curved) simplifies the deduction of the instantaneous heat flux from the measure surface temperature history, it can lead to substantial inaccuracy when spatial variations of surface heat flux are present, or when the test object has strong curvature. George and Reinecke \cite{George1963} analysed errors arising due to lateral conduction within thin-skinned test objects operating under calorimeter principles. The error in the one dimensional analysis due to lateral conduction effects initially increases in a linear manner with time in transient experiments \cite{George1963}. Using a similar analysis, Schultz and Jones \cite{Schultz1973} illustrate a particular example of lateral conduction: a 20\,mm diameter nickel hemisphere in hypersonic flow will register a 5\,\% error in stagnation point heat flux after 130\,ms if lateral conduction effects are ignored.   

\subsubsection{Lateral Conduction Corrections}

To correct for lateral conduction effects in the case of thin film transient heat flux measurements near the stagnation point on a hemispherical-nosed cylinder of radius $R$, the stagnation point heat flux $q_0$ can be decomposed to the normal and lateral components within the substrate
\begin{equation}
    q_0 = q_n + q_l
    \label{eq:qs_is_qn_plus_ql}
\end{equation}
and an approximate expression is developed in  \cite{Buttsworth1998b} for the lateral conduction which is given as 
\begin{equation}
    q_l(t) = \frac{2\alpha}{R^2} \int_0^t \left( \frac{\partial^2 q}{\partial\theta^2} \right)_{\theta=0} d\tau
\end{equation}
where $\alpha$ is the thermal diffusivity of the probe substrate (a constant value), and $\theta$ is the polar co-ordinate with $\theta = 0$ defining the location of the stagnation point. The surface heat flux was then approximated as 
\begin{equation}
    q(t,\theta) = q_0(t) \, g(\theta) ,
\end{equation}
where the function $g(\theta)$ is determined from separate theoretical or empirical results.  The expression for the lateral conduction can then be written as
\begin{equation}
    q_l(t) = \frac{2\alpha}{R^2} \left( \frac{d^2 g}{d \theta^2} \right)_{\theta=0}\int_0^t q_0 d\tau .
    \label{eq:lateral_heat_jturbomodel}
\end{equation}
In practice, the normal component of heat flux $q_n$ is first determined from the surface temperature history using a one-dimensional model for the transient heat conduction within the substrate (while accommodating the radial geometry of the hemisphere as necessary), and the stagnation point heat flux in (\ref{eq:lateral_heat_jturbomodel}) is approximated as $q_0 \approx q_n$.  The lateral conduction term is then calculated from (\ref{eq:lateral_heat_jturbomodel}) and an improved approximation for the stagnation point heat flux is calculated from (\ref{eq:qs_is_qn_plus_ql}), and iteration continues until convergence of $q_l(t)$ is achieved.

In the case of thermochromic liquid crystal experiments where the heat transfer coefficient is deduced from the recorded time at which the colour change occurs, Kingsley-Rowe et al. \cite{KingsleyRowe2005} considered the ratio of lateral to normal conduction as a constant value to enable solution of the two-dimensional transient heat conduction equation as a modified form of the one-dimensional problem.  Although the ratio of the lateral-to-normal conduction will actually change with time, the task in \cite{KingsleyRowe2005} was to select an appropriate average value for which the lateral conduction error is minimised, and this was achieved through finite difference computations with representative spatial distributions of heat transfer coefficient.  The method of \cite{KingsleyRowe2005} was subsequently extended by Brack et al. \cite{Brack2016} to three-dimensional cases, where lateral conduction is two-dimensional.  

\subsubsection{Direct, Inverse and Iterative Methods}

Corrections for lateral conduction discussed in the previous section were based on approximations for the multi-dimensional transient heat conduction equation that enabled the application of the one-dimensional semi-infinite heat conduction solution in a modified manner.  We now consider previous approaches which in which surface temperature measurements with multi-dimensional transient heat conduction effects have been analysed without recourse to the one-dimensional approximation.  

Direct approaches involve the application of the measured surface temperature history as a Dirichlet boundary condition in a model for the substrate heat conduction in order to deduce the surface heat flux.  Solano and Paniuagua \cite{Solano2009} used a direct approach based on a two-dimensional finite element model of their rotor blade at mid-height. The surface heat flux was then determined from the history of the temperature distribution around the perimeter of rotor, which was reconstructed from a modest number of thin film measurements \cite{Solano2009}.  Ling et al. \cite{Ling2004} also used a direct approach based on a three-dimensional finite difference scheme and used thermochromic liquid crystal for surface temperature measurement.  The direct method was possible because of the high spatial resolution in the measured surface temperature history which was applied directly as the Dirichlet boundary condition in the finite difference calculations \cite{Ling2004}.

Inverse techniques are typically required in cases where the temperature measurement location does not coincide with the physical position at which the surface heat flux is required.  Walker et al. \cite{Walker2000} demonstrated the application of an inverse technique for deduction of heat flux from multiple thin film temperature sensors in a shock interaction experiment which produced strong lateral gradients of temperature and heat flux.  Although the thin film gauges were on the surface where the heat flux was required, the inverse approach was preferred because of the finite number of temperature sensors and associated uncertainties in reconstruction of the surface temperature history \cite{Walker2000}. Sousa et al. \cite{Sousa2012} used a finite element solver in an iterative manner to treat a transient inverse problem in which temperatures were measured using IR cameras focussed on a surface of the domain which was different from the heat flux surface of primary interest.

Iterative methods commence with an initial guess for the distribution and history of the surface heat flux (or surface heat transfer coefficient) and the resulting surface temperature history is then calculated using a model for the substrate heat conduction. The surface flux is then revised based on the difference between the simulated and measured surface temperature history and iterations continue until the simulated temperatures match the measured values with sufficient accuracy.  Lin and Wang \cite{Lin2002} used such an approach and developed a three dimensional finite difference scheme for modelling the substrate heat conduction associated with an impinging jet configuration with surface temperature measurements achieved using thermochromic liquid crystal. In analysing their experimental data, Lin and Wang \cite{Lin2002} demonstrated differences in deduced heat flux values of up to 20\,\% between the standard one-dimensional analysis and their new three-dimensional analysis.  Ryley et al. \cite{Ryley2019} also applied a similar approach using a finite element analysis of a considerably more complex geometry to determine the distribution of the heat transfer coefficient across the surface.

\subsection{Summary}
Previous work in the analysis of transient heat transfer experiments where multi-dimensional heat conduction effects are significant has demonstrated a wide variety of techniques ranging from approximate corrections to one dimensional analytical results through to iterative approaches using high-fidelity, finite element models of complex substrates.
The computational simplicity offered by analytical results is appealing, but it may not be possible to achieve high accuracy from previous analytical techniques in the deduced heat flux values when the substrate includes fine geometric features or when the experiment needs to be performed for an extended period of time.
One analytical approach that has potential for application in multi-dimensional substrate heat conduction problems involves heat kernel methods.
The purpose of this work to demonstrate the application of Neumann heat kernel results in the analysis of transient heat transfer experiments with multi-dimensional heat conduction effects within the substrate.  

\section{Motivating Example: The Hemispherical Probe}
\label{sec:MotivatingExample}

\subsection{Hardware and Background}

\begin{figure}
 \centering
 \includegraphics[width=89mm]{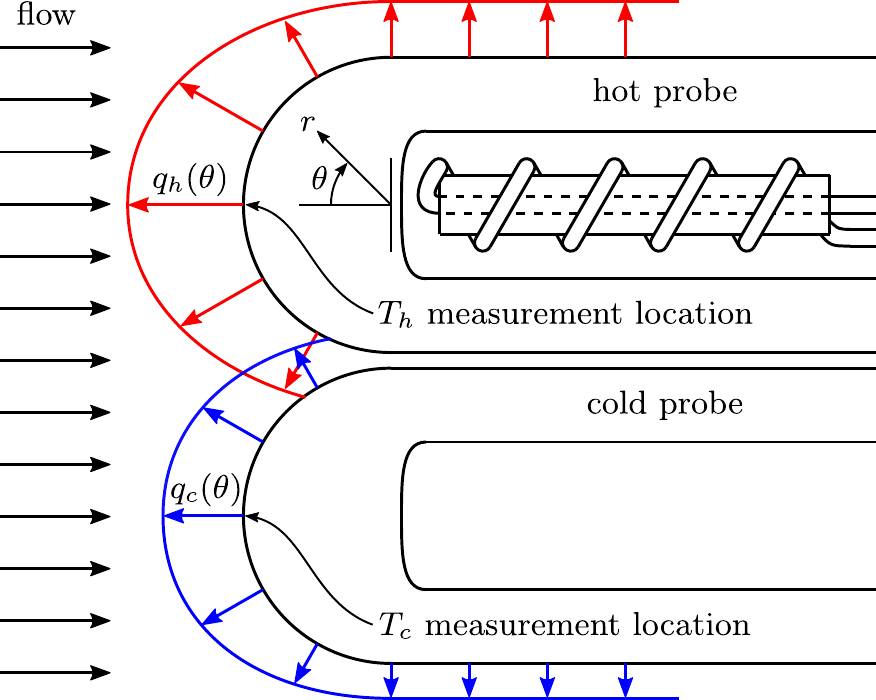}
 \caption{Temperature probe based on twin thin film heat flux gauges operated at different temperatures as used in \cite{Buttsworth1998c}.}
 \label{fig:pyestock_probe}
\end{figure}

The physical device considered in the present work is illustrated in Fig.~\ref{fig:pyestock_probe}.  The device involved two nominally identical cylinders with an external diameter of 3\,mm and with approximately hemispherical tips.  The probes were made of fused silica with platinum thin film temperature sensors located close to the stagnation point on each probe.  The probe arrangement was operated downstream of an experimental high pressure turbine as described in \cite{Buttsworth1998c}.  One of the probes included an internal electrical heating element so that the two probes would operate at different surface temperatures.  Having the probes at two different surface temperatures enables deduction of the flow stagnation temperature on the assumption that the convective heat transfer coefficient for each probe is the same.  

\begin{figure}
 \centering
 \includegraphics[width=89mm]{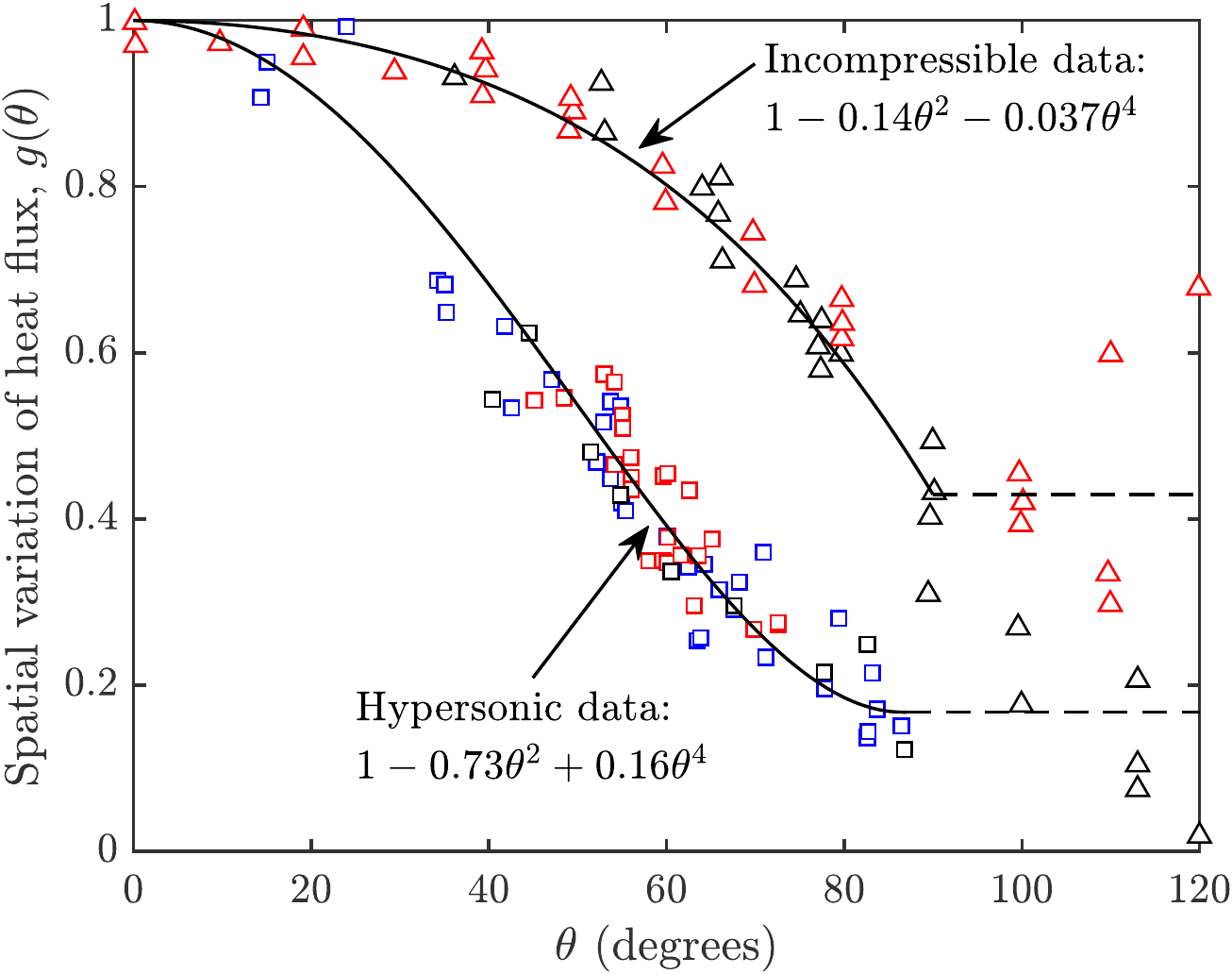}
 \caption{Spatial variation of heat flux around a hemispherical-nosed cylinder in hypersonic flow (data from \cite{Kemp1959}) and around a sphere in incompressible flow (data from compilation by \cite{Galloway1968}).}
 \label{fig:sphere_data}
\end{figure}

The flow duration of each turbine experiment was $\sim 0.5$\,s, and over such a period, it was not possible to treat the substrate heat conduction using a flat-surface semi-infinite approximation.  This is demonstrated by considering the thermal diffusivity of the fused silica substrate which was $\alpha \sim 8\times10^{-7}$\,m$^2$/s giving a value of $\sqrt{\alpha t} \sim 0.6$\,mm and it is therefore recognised that heat will have penetrated a substantial distance relative to the outer radius of the probe which is $R = 1.5$\,mm.  Furthermore, each probe will experience a spatial distribution of heat flux around its perimeter and because of the relatively long duration of the experiment, significant lateral conduction effects can also be expected.  The spatial variation of heat flux around the windward surface of spheres and hemispherical-nosed cylinders is illustrated in Fig.~\ref{fig:sphere_data}. The spatial variation is more gentle for incompressible flow, but even in this case, the heat flux at a polar angle of $\theta = 90^\circ$ is less than half that at the stagnation point (see Fig.~\ref{fig:sphere_data}).

Errors in the modelling of the transient heat conduction within the substrate that may arise due to radial and lateral conduction effects will propagate through the analysis process and potentially corrupt the deduction of flow total temperature.  Therefore, a separate experiment involving an atmospheric blow-down duct was established to enable independent verification of the substrate heat conduction modelling. The hardware associated with the verification blow-down experiment is described elsewhere \cite{Buttsworth1998c}, but the same probes are used.

\subsection{Results}

\begin{figure}
 \centering
 \includegraphics[width=89mm]{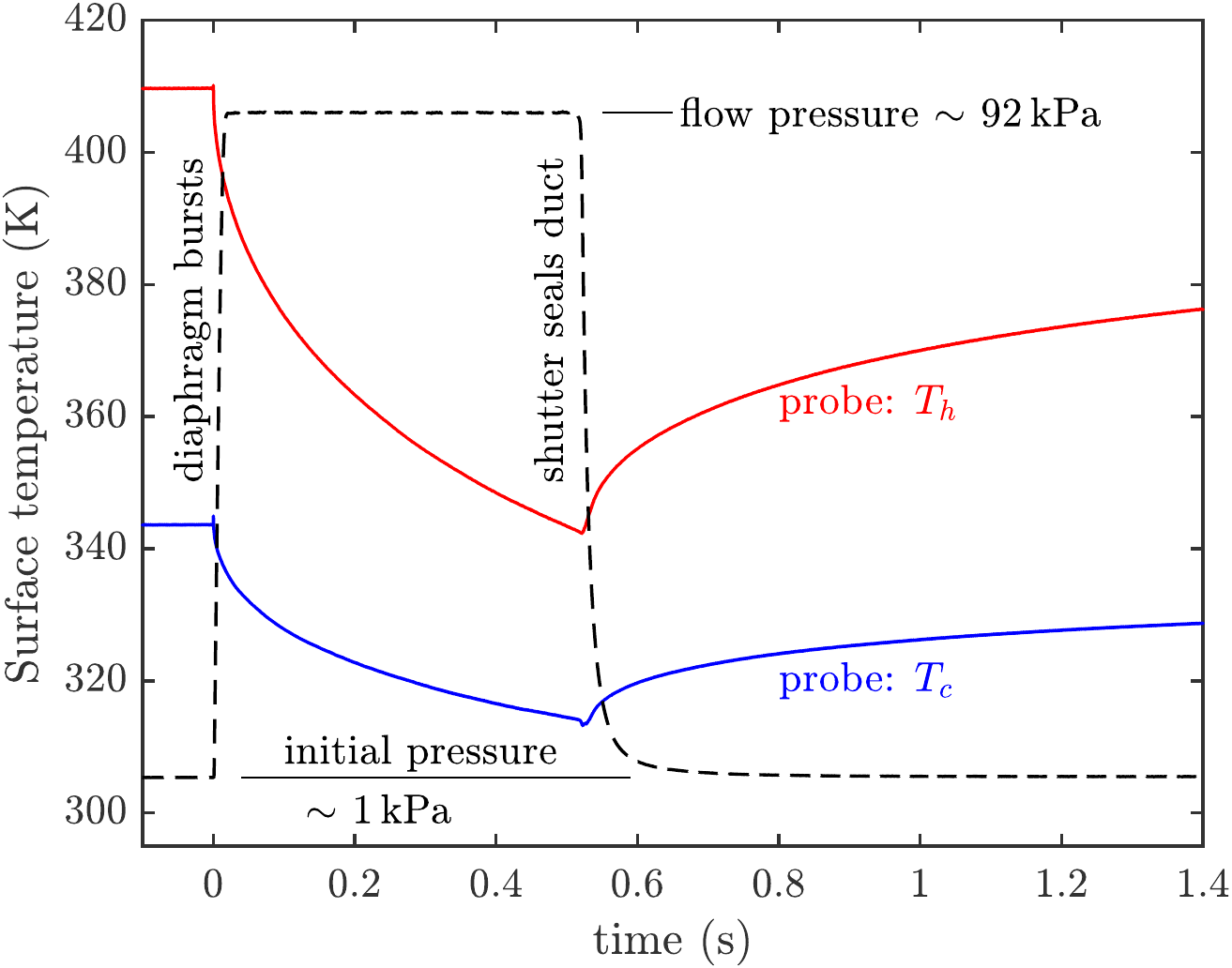}
 \caption{Surface temperature histories for the hot and cold probes, and the static pressure history within the duct.}
 \label{fig:p_and_T_exp}
\end{figure}

Results from the verification blow-down experiment are illustrated in Fig.~\ref{fig:p_and_T_exp}. The duct was connected to a large vacuum tank which was evacuated to an  initial absolute pressure of around 1\,kPa.  To initiate an experiment, the diaphragm at the bell-mouth entrance to the duct was ruptured and this event occurs at $t = 0$ in Fig.~\ref{fig:p_and_T_exp}. Air from the ambient environment is drawn into the duct and steady flow conditions are rapidly established within the duct, as reflected in the essentially constant duct static pressure measurement which gives a value of $\sim 92$\,kPa.  Based on such static pressure measurements, the flow Mach number in the vicinity of the probe tips was found to be $M_\infty \sim 0.4$ \cite{Buttsworth1998c}, and under such conditions it is expected that the correlation given by
\begin{equation} \label{eq:gfunction_incomp}
    g(\theta) = \frac{q}{q_0} = 1 - 0.14 \theta^2 - 0.037 \theta^4
\end{equation}
over the range $0 \le \theta \le \frac{\pi}{2}$ will provide  a reasonable approximation for the spatial variation of flux, as illustrated in Fig.~\ref{fig:sphere_data}. 

The flow stagnation temperature in this experiment was equal to the ambient laboratory temperature, which was measured as 293.0\,K on this occasion.  Since the initial temperature of the hot and cold probes was higher than the flow stagnation temperature, the measured probe temperatures decrease with the onset of flow within the duct, Fig.~\ref{fig:p_and_T_exp}.  Rapid termination of the flow  was achieved at about 0.52\,s by a shutter that was arranged to slam onto the duct entrance.  At the time of duct closure, the pressure rapidly falls to a value close to that of the initial pressure.  Therefore, the suitability of the transient heat flux model that is used to analyse the probe surface temperature measurements can be assessed by examining the values of apparent heat flux after flow termination, which should essentially be zero for both probes. 

\subsection{Limitations of Earlier Heat Flux Analyses}

\begin{figure}
 \centering
 \includegraphics[width=89mm]{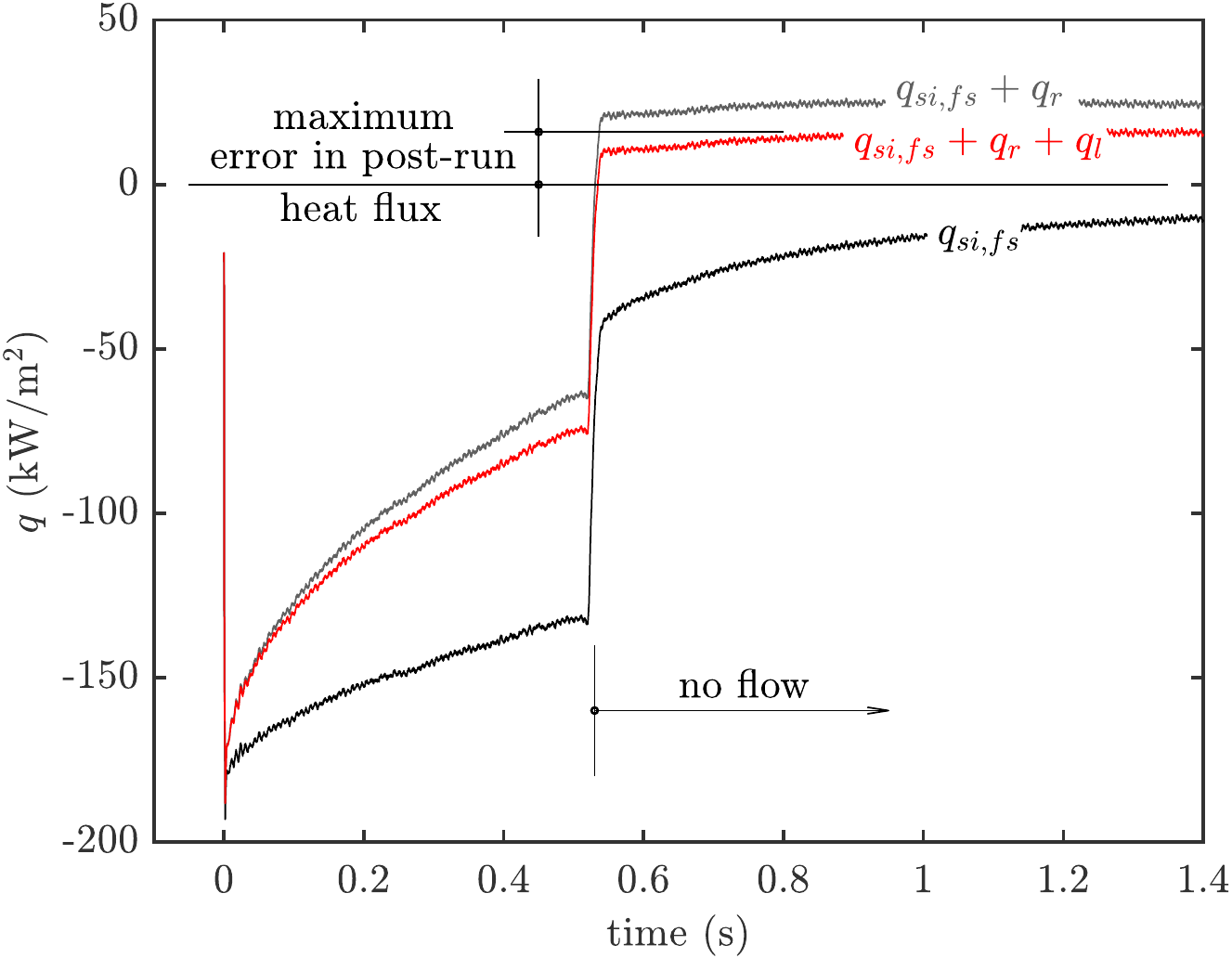}
 \caption{Heat flux results for the hot probe based on the semi-infinite analysis with corrections for radial conduction and lateral conduction effects.}
 \label{fig:heatflux_exp_1}
\end{figure}

The former approach used in the deduction of heat flux involved application of a finite difference scheme which simultaneously accommodated both the sphere-like substrate geometry and temperature-dependent thermal properties \cite{Buttsworth1998c}. However, as the finite difference method was only one-dimensional, a correction for the lateral conduction effects was still required.    Figure~\ref{fig:heatflux_exp_1} illustrates the magnitudes of the different effects in the case of the heated probe, without recourse to the finite difference method used previously. The result labelled $q_{si,fs}$ in Fig.~\ref{fig:heatflux_exp_1} was obtained by treating the $T_h$ result in Fig.~\ref{fig:p_and_T_exp} with a semi-infinite flat surface analysis.  The correction for curvature effects given by \eqref{eq:q_r} was then applied, giving the corrected result labelled $q_{si,fs} + q_r$.  Finally, the lateral conduction effects were then calculated using \eqref{eq:lateral_heat_jturbomodel}, giving the corrected result labelled $q_{si,fs} + q_r + q_l$ in Fig.~\ref{fig:heatflux_exp_1}, but it is noted that there remains a non-zero value of heat flux after termination of the flow, in the period for $t > 0.55$\,s in Fig.~\ref{fig:heatflux_exp_1}.

\begin{figure}
 \centering
 \includegraphics[width=89mm]{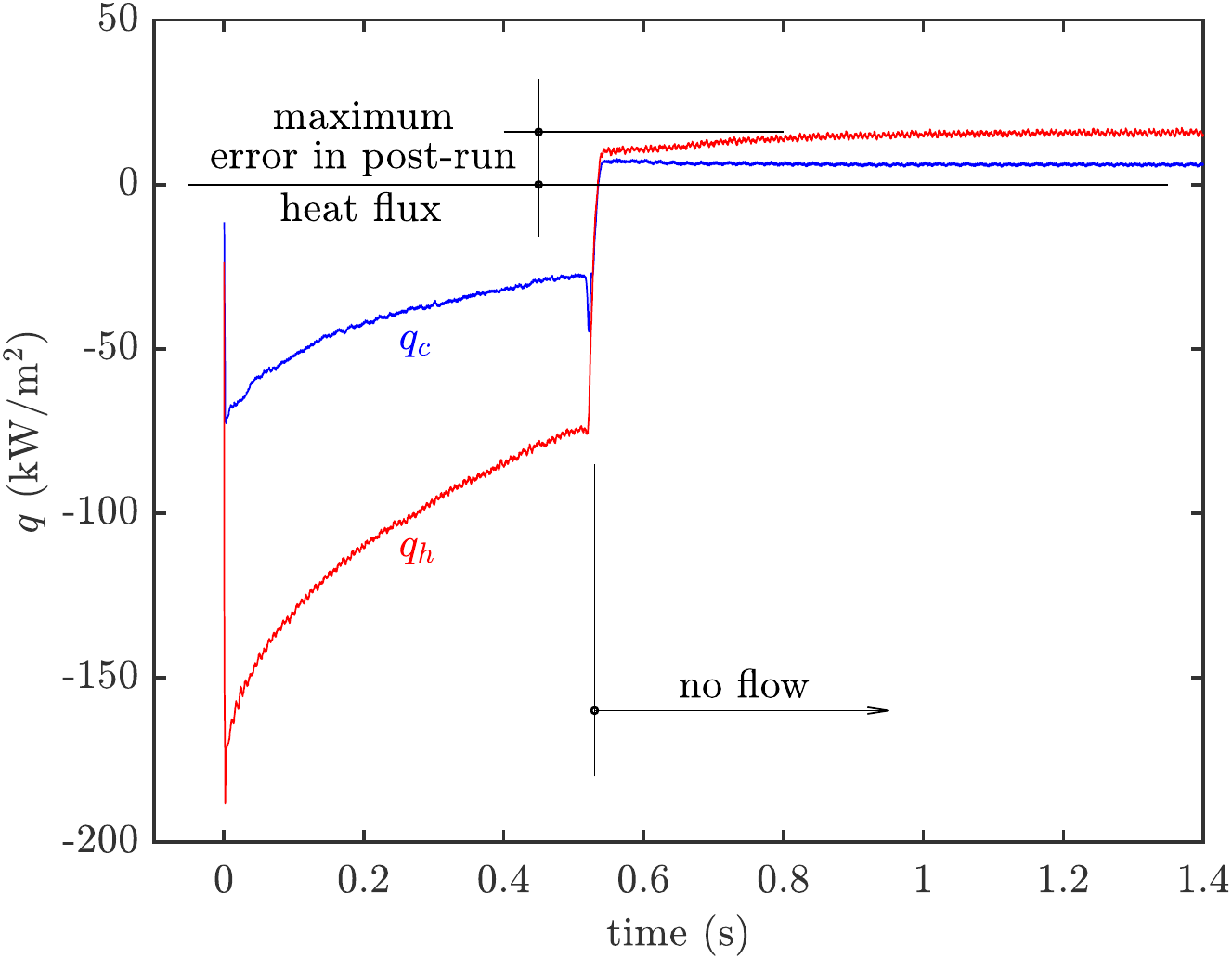}
 \caption{Heat flux results for the hot probe and the cold probe using corrections to the semi-infinite flat surface analysis method.}
 \label{fig:heatflux_exp_2}
\end{figure}

Figure~\ref{fig:heatflux_exp_2} re-presents the Fig.~\ref{fig:heatflux_exp_1} result $q_{si,fs} + q_r + q_l$ with the label $q_h$, for the purpose of comparison with the cold probe $T_c$ result labelled $q_c$ in Fig.~\ref{fig:heatflux_exp_2}.  Both the $q_h$ and the $q_c$ results in Fig.~\ref{fig:heatflux_exp_2} were obtained in the same manner and significant non-zero heat flux values after flow termination for $t > 0.55$\,s are observed in both cases.
Variable thermal properties within the probe substrates will have an effect, and more so in the case of the hot probe since its surface temperature decreased by 67.4$^\circ$C during the flow time whereas that of the cold probe decreased by only 29.6$^\circ$C.  Errors due to variable thermal property effects were assessed using the finite difference methods applied the original work \cite{Buttsworth1998c}. However, the magnitude of the error due to variable property effects in the post-flow heat flux is about 1.8\,kW/m$^2$ in the hot probe case and about 0.55\,kW/m$^2$ in the cold probe case, which cannot fully explain the magnitude of the errors observed in Fig.~\ref{fig:heatflux_exp_2}.

In the earlier work \cite{Buttsworth1998c}, errors in the post-flow heat flux were minimised by selecting values of $\left( \frac{d^2 g}{d \theta^2} \right)_{\theta=0}$ so that the immediate post-flow values of heat flux for times $0.55 \lesssim t \lesssim 0.62$ were approximately zero.  In so doing, the required magnitude of $\left( \frac{d^2 g}{d \theta^2} \right)_{\theta=0}$ was unrealistically large: in the case of the hot probe a value of about $-0.53$ was used, and in the case of the cold probe a value of about $-0.64$ was used.  The magnitude of these values is around twice as large as that from the incompressible data in Fig.~\ref{fig:sphere_data} and \eqref{eq:gfunction_incomp} which suggests 
\begin{equation}
    \left( \frac{d^2 g}{d \theta^2} \right)_{\theta=0} = -0.28.
\end{equation}
Furthermore, the fact that different values were required for the hot and the cold probe highlights the empirical nature of the former approach which is not necessarily justified by other observations.  

\subsection{Heat Equation}
The tips of the probes are approximately hemispherical, and the period of the experiment is such that significant corrections to the semi-infinite flat surface analysis are required. To develop a higher fidelity analysis method, the heat equation that accommodates both radial and lateral conduction effects should be used:     
\begin{equation}
    \frac{1}{\alpha} \frac{\partial T}{\partial t} = \frac{\partial^2 T}{\partial r^2} + \frac{2}{r} \frac{\partial T}{\partial r} + \frac{2}{r^2} \frac{\partial^2 T}{\partial \theta^2}+\frac{\cot(\theta)}{r^2}\frac{\partial T}{\partial \theta} .
\end{equation}
In specifying the above model for the heat conduction, the flow around each probe is assumed to be symmetric about its own stagnation streamline. The thermal properties are assumed to be constant; they are incorporated into a single parameter, the thermal diffusivity, which is given by
\begin{equation}
    \alpha = \frac{k}{\rho c_p} ,
\end{equation}
were $k$ is the thermal conductivity, $\rho$ is the density, and $c_p$ is the specific heat of the substrate material. The heat flux at the surface of the substrate is given by
\begin{equation}
    q = k \left( \frac{\partial T}{\partial r} \right)_{r=R},
\end{equation}
so that a flux of heat into the surface of the substrate is a positive quantity.

It is convenient to introduce the mathematics of the higher fidelity analysis in non-dimensional form, so we can consider a non-dimensional radius given by
\begin{equation}
    \ndv{r} = \frac{r}{R} ,
\end{equation}
where $R$ is the radius of the hemisphere, and a non-dimensional time given by
\begin{equation}
    \ndv{t} = \frac{\alpha t }{R^2} ,
\end{equation}
and a non-dimensional temperature given by
\begin{equation}
    u  = \frac{T}{T_i} - 1 ,
\end{equation}
where $T_i$ is the initial temperature of the substrate, taken to be a constant value.  The heat equation in the present application is then 
\begin{equation}
    \frac{\partial u}{\partial \ndv{t}} = \frac{\partial^2 u}{\partial \ndv{r}^2} + \frac{2}{\ndv{r}} \frac{\partial u}{\partial \ndv{r}} + \frac{2}{\ndv{r}^2} \frac{\partial^2 u}{\partial \theta^2}+\frac{\cot(\theta)}{\ndv{r}^2}\frac{\partial u}{\partial \theta} ,
\end{equation}
which, in more compact notation can be written as
\begin{equation}
\frac{\partial u}{\partial \ndv{t}} = \Delta u  
\end{equation}
and the non-dimensional heat flux at the surface of the substrate is given by
\begin{equation}
    c = q \frac{R}{k \, T_i} .
\end{equation}

\subsection{Approach}

\begin{figure}
 \centering
 \includegraphics[width=89mm]{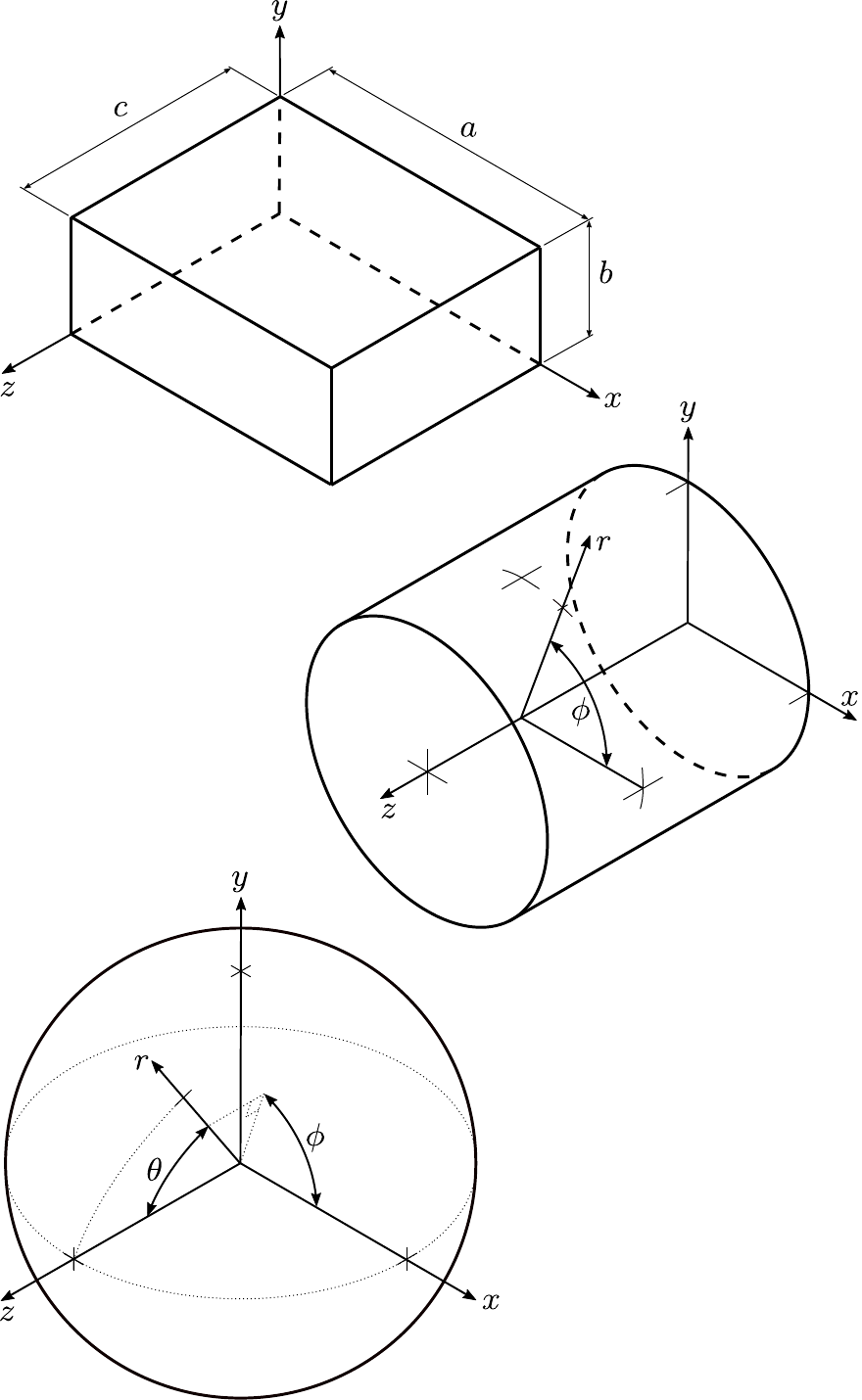}
 \caption{Illustration of geometries for which Neumann heat kernel results are developed.}
 \label{fig:box_cylinder_ball}
\end{figure}

\begin{figure}
 \centering
 \includegraphics[width=89mm]{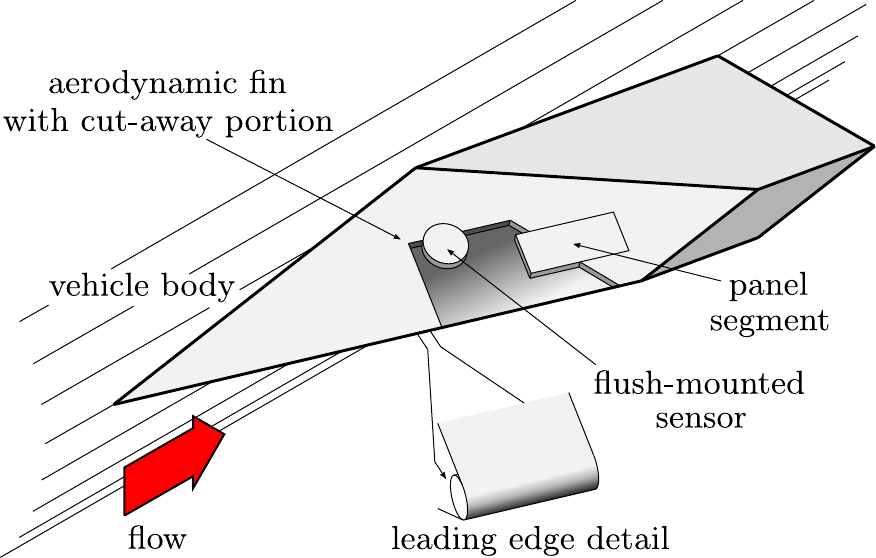}
 \caption{Potential applications for the box and cylinder heat kernel results in the context of experiments with an aerodynamic fin on a body.}
 \label{fig:HK_applications}
\end{figure}

In the experiments, temperature was effectively measured at a single location -- the stagnation point -- and the relative distribution of the heat flux around the perimeter is known with reasonable precision from separate experiments.
Instead of correcting the semi-infinite flat surface analysis for both radial and lateral conduction effects, we have developed a method whereby these effects can be simultaneously accommodated in a single step.  To achieve this single-step analysis, the Neumann heat kernel, which will be described in Section~\ref{MOHKS}, is used and specific results are developed for 3 different geometries: the solid box, the solid cylinder and the solid sphere, as illustrated in Fig.~\ref{fig:box_cylinder_ball}.  Such geometries provide higher fidelity models for substrates than offered by the semi-infinite flat surface model.  The application of the sphere heat kernel result to the analysis of the hemispherical-nosed probe is then detailed in Section~\ref{sec:Applications}. While applications for the other two configurations -- the cylinder and the box -- are not described in this paper, we expect several opportunities for such applications exist.  For example, the box heat kernel may prove useful in the analysis of convective heat flux on a rectangular panel or other box-like structures, and the cylinder heat kernel may find application in the analysis of either leading edges of aerodynamic structures or disk-like sensors, as illustrated in Fig.~\ref{fig:HK_applications}.

\section{Mathematics}\label{MOHKS}

For the remainder of this paper, the $\ndv{(\hspace{2mm})}$ notation is dropped from the spatial variables $r$, $x$, and $y$, although it is retained to distinguish the non-dimensional time $\ndv{t}$ from the dimensional time $t$.

\subsection{The Mathematical Statement of the Problem}
Let $\Omega\subseteq \mathbb{R}^n$ be our domain with boundary $\partial \Omega$. 
Mathematically, the problem we are faced with is solving the non-dimensional heat equation 
\begin{equation}\label{HE}
\frac{\partial u}{\partial \ndv{t}}(\ndv{t},x)=\Delta u(\ndv{t},x), \  (\ndv{t},x)\in (0,\infty)\times \Omega.  
\end{equation}
For simplicity, we impose zero initial conditions:
\begin{equation}\label{IC}
u(0,x)=0.
\end{equation}
Since we know the spatial variation of heat flux $g(x)$ across $\partial \Omega$, we have the following Neumann conditions:
\begin{equation}\label{NC'}
\nabla u(\ndv{t},x)\cdot \nu(x)=c(\ndv{t})g(x), \ (\ndv{t},x)\in (0,\infty)\times \partial \Omega,
\end{equation}
where $\nu(x)$ is the outward pointing normal vector at $x\in \partial \Omega$, and $c(\ndv{t})$ is \textit{a priori} unknown. In order to determine $c(\ndv{t})$, we measure Dirichlet conditions at a convenient point $x_0\in \partial \Omega$:
\begin{equation}\label{DC}
u(\ndv{t},x_0)=d(\ndv{t}), \ \ndv{t}\in (0,\infty).
\end{equation}
For compatibility, we impose that $c(0)=d(0)=0$.  The problem is then as follows. \\

\noindent
\textbf{Problem.}
Fix a flux function $g(x)$ and a measurement function $d(\ndv{t})$. Find functions $u:(0,\infty)\times \Omega\to \mathbb{R}$ and $c:(0,\infty)\to \mathbb{R}$ satisfying \eqref{HE}-\eqref{DC}. 

\subsection{The Solution: The Utility of the Neumann Heat Kernel}\label{UNHK}
Since Neumann conditions are the dominating component of our boundary conditions in the Problem, we briefly discuss the Neumann heat problem, i.e., the problem of solving \eqref{HE} for $u(\ndv{t},x)$ subject to 
\begin{equation}\label{NC}
\nabla u(\ndv t,x)\cdot \nu(x)=0, \ (\ndv{t},x)\in (0,\infty)\times \partial \Omega,
\end{equation}
and arbitrary initial conditions. One of the most powerful tools for the systematic treatment of this problem is the so-called \textit{Neumann heat kernel}, a.k.a. the fundamental solution to the Neumann heat problem. \\

\noindent
\textbf{Definition.}
On a given domain $\Omega\subset \mathbb{R}^n$ with boundary $\partial \Omega$, a Neumann heat kernel 
is a function $p:(0,\infty)\times \Omega\times \Omega\to \mathbb{R}$ satisfying the following:
\begin{enumerate}
  \item $\frac{\partial p(\ndv{t},x,y)}{\partial \ndv{t}}=\Delta_y p(\ndv{t},x,y)$ ($p$ satisfies the heat equation)
  \item For all $y\in \partial \Omega$, we have $\nabla_y p(\ndv{t},x,y)\cdot \nu(y)=0$, 
  ($p$ satisfies Neumann conditions).
  \item If $f:\Omega\to \mathbb{R}$ is smooth, then $\lim_{\ndv{t}\to 0}\int_{\Omega}p(\ndv{t},x,y)f(y)dy=f(x)$ ($p$ satisfies the initial conditions $p(0,x,y)=\delta(x-y)$, 
  where $\delta$ is the Dirac delta distribution). 
\end{enumerate}
Thus, a Neumann heat kernel is a solution to the heat equation with Neumann conditions, and initial conditions coinciding with a total unit energy compactified into a single point. \\

The Neumann heat kernel can be used to construct a solution of the heat equation with \textit{non-zero} Neumann conditions. Indeed, in equation \eqref{HE}, we replace $\ndv{t}$ and $x$ by $\tau$ and $y$ respectively, multiply  by $p(\ndv{t}-\tau,x,y)$ and integrate over $\Omega$ and $[0,\ndv{t}-\epsilon]$:
\begin{align*}
    \int_{0}^{\ndv{t}-\epsilon}\int_{\Omega}\frac{\partial u}{\partial \tau}(\tau,y)p(\ndv{t}-\tau,x,y)dyd\tau=\int_{0}^{\ndv{t}-\epsilon}\int_{\Omega}\Delta_y u(\tau,y)p(\ndv{t}-\tau,x,y)dyd\tau. 
\end{align*}
Using integration by parts, and the initial conditions \eqref{IC} for $u$, we find 
\begin{align*}
    \int_{\Omega}u(\ndv{t}-\epsilon,y)p(\epsilon,x,y)+\int_{0}^{\ndv{t}-\epsilon}\int_{\Omega}u(\tau,y)\frac{\partial p}{\partial \ndv{t}}(\ndv{t}-\tau,x,y)dyd\tau\\=\int_{0}^{\ndv{t}-\epsilon}\int_{\Omega}u(\tau,y)\Delta_y p(\ndv{t}-\tau,x,y)dyd\tau+\int_{0}^{\ndv{t}-\epsilon}\int_{\partial \Omega}p(\ndv{t}-\tau,x,y)\nabla_y u(\tau,x,y)\cdot \nu(y)dy.
\end{align*}
Since $p$ solves the heat equation, and $u$ satisfies \eqref{NC'}, we find that 
\begin{align*}
\int_{\Omega}u(\ndv{t}-\epsilon,y)p(\epsilon,x,y)dy    &=\int_{0}^{\ndv{t}-\epsilon}\int_{\partial \Omega}p(\ndv{t}-\tau,x,y)c(\tau)g(y)dyd\tau.
\end{align*}
By sending $\epsilon$ to $0$ and using the initial conditions for $p$, we find \\

\noindent
\textbf{Theorem.}
\textit{The solution of \eqref{HE}-\eqref{NC'} is given by the convolution integral}
\begin{align}\label{ME}
 u(\ndv{t},x)&=\int_{0}^{\ndv{t}}c(\tau)\int_{\partial \Omega}p(\ndv{t}-\tau,x,y) g(y)dy d\tau
\end{align}

We can use \eqref{ME} with $x=x_0$ and \eqref{DC} to find that 
\begin{equation}\label{EFLT}
d(\ndv{t})=\int_{0}^{\ndv{t}}c(\tau)\int_{\partial \Omega}p(\ndv{t}-\tau,x_0,y) g(y)dy d\tau.
\end{equation}
We refer to the quantity
\begin{equation}\label{IRF}
 P_g(\ndv{t}) : = \int_{\partial \Omega}p(\ndv{t},x_0,y) g(y) dy
\end{equation}
as the \textit{Impulse Response Function}. In this notation, \eqref{EFLT} becomes 
\begin{equation}\label{IRFconv}
d(\ndv{t})=\int_{0}^{\ndv{t}}c(\tau)P_g(\ndv{t}-\tau)d\tau.
\end{equation}
Since $d(\ndv{t})$ is known, we can recover $c(\ndv{t})$ from \eqref{IRFconv} using Laplace transforms, and then use \eqref{ME} to find $u(\ndv{t},x)$. This completes the solution to the Problem, provided we know a Neumann heat kernel for $\Omega$. \\

\begin{figure}
 \centering
 \includegraphics[width=89mm]{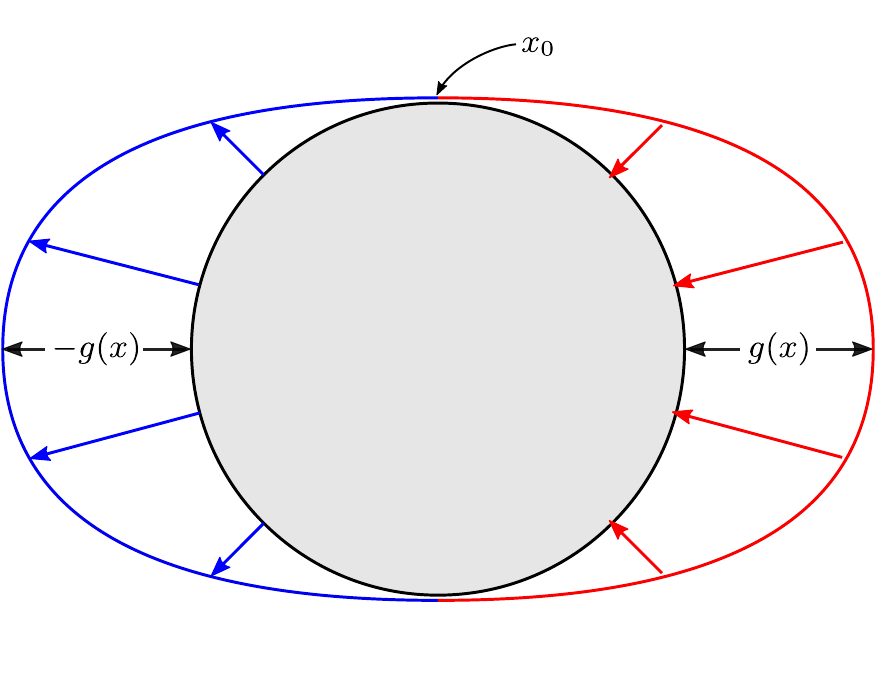}
 \caption{Illustration of a problematic choice of $x_0$.}
 \label{fig:OddFunction}
\end{figure}

\noindent
\textit{Remark.}
For a given $d(\ndv{t})$, solving \eqref{EFLT} for $c(\ndv{t})$ is not always possible. Indeed, if $g=0$, then $d$ must obviously be uniformly $0$. For a less trivial example, take $\Omega$ to be the two-dimensional sphere (so $\partial \Omega$ is the circle), $x_0$ to be the top of the circle, $g$ to be any function which is odd about the vertical axis, as illustrated in Fig.~\ref{fig:OddFunction}. Then $d$ must still be $0$ uniformly.  If for example $g$ is positive on the entire boundary, then the impulse response function is strictly positive and we can recover $c(\ndv{t})$ by using Laplace transforms on \eqref{IRFconv}. 

\subsection{Constructing the Neumann Heat Kernel}
In Section \ref{UNHK}, we discussed the definition of a Neumann heat kernel. We subsequently solved \eqref{HE}-\eqref{DC} in terms of this heat kernel. We now turn to the general problem of constructing the heat kernel for a given domain $\Omega$. We begin our discussion with some simple examples.  \\

\noindent
\textbf{Example: Infinite Rod} \\
The simplest heat kernel arises when $\Omega=\mathbb{R}$. In this situation, $\Omega$ has no boundary, so Neumann conditions have no meaning. It is well known that the heat kernel (fundamental solution) is given by 
\begin{align*}
    p(\ndv{t},x_1,x_2)=\frac{1}{\sqrt{4\pi \ndv{t}}}e^{-\frac{(x_1-x_2)^2}{4\ndv{t}}}.
\end{align*}

\noindent
\textbf{Example: Semi-infinite Rod} \\
We now turn to $\Omega=[0,\infty)$, a domain which actually has a non-empty boundary of $\{0\}$. The Neumann heat kernel can be found by altering the one in the Infinite Rod example:
\begin{align*}
   p(\ndv{t},x_1,x_2)=\frac{1}{\sqrt{4\pi \ndv{t}}}\left(e^{-\frac{(x_1-x_2)^2}{4\ndv{t}}}+e^{-\frac{(x_1+x_2)^2}{4\ndv{t}}} \right).
\end{align*}
Note that when $g(y) = 1$, the impulse response function \eqref{IRF} for this semi-infinite example becomes
\begin{align*}
    P_{g,si,fs}(\ndv{t}) = \frac{1}{\sqrt{\pi \ndv{t} }} ,
\end{align*}
which is a fundamental result that is particularly relevant to the transient heat transfer measurement technique for substrates that are treated as semi-infinite flat surfaces. \\

The domains in these two examples are both non-compact manifolds, but most physically-relevant manifolds are compact, with boundary. For such manifolds, it is essentially impossible to have heat kernels with such nice closed-form representations, but we can still make useful approximations. Indeed, if $\Omega$ is a compact manifold with boundary $\partial \Omega$, then we can write 
\begin{align}\label{HKWEF}
    p(\ndv{t},x,y)=\sum_{n=1}^{\infty}e^{-\lambda_n \ndv{t}}\psi_n(x)\psi_n(y),
\end{align}
where $\lambda_n$ are the eigenvalues of the Laplacian with Neumann conditions, and $\psi_n$ are the corresponding $L^2(\Omega)$-orthonormal eigenfunctions, i.e.,
we have 
\begin{align*}
    \Delta \psi_n(x)+\lambda_n \psi(x)&=0, \ x\in \Omega,\\
    \nabla \psi_n(x)\cdot \nu(x)&=0, \ x\in \partial \Omega,\\
    \int_{\Omega} \psi_n^2(x)dx&=1,\\
    \int_{\Omega} \psi_n(x)\psi_m(x)dx&=0, \ \text{if} \ n\neq m.
\end{align*}
Therefore, for large $N\in \mathbb{N}$, 
the expression 
\begin{align*}
    \sum_{n=1}^{N}e^{-\lambda_n \ndv{t}}\psi_n(x)\psi_n(y)
\end{align*}
\textit{should} be an approximation for $p$. Verifying that this is true in general requires knowledge of the mathematics of PDEs, specifically, elliptic regularity theory, Sobolev embedding theorems and Weyl's law (see, for example, Chapters 5 and 6 of \cite{Evans}). 
There are other methods available to approximate the heat kernel (see, for example, Chapter 23 of \cite{Chow}), but in this paper, we will focus on geometrically simple domains where we know the eigenvalues and eigenfunctions, and can therefore approximate the heat kernel using \eqref{HKWEF}. These eigenfunctions and eigenvalues are found using the well-known method of \textit{Separation of Variables}, as described, for example, in Chapter 12 of \cite{Kreyszig}. 
\\

\noindent
\textbf{Example: The Box} \\
In this example, we set $\Omega=\{(x,y,z): 0\le x\le a, 0\le y\le b, 0\le z\le c\}$. The heat kernel is given by 
\begin{align*}
    p(\ndv{t},(x_1,y_1,z_1),(x_2,y_2,z_2))
    =\sum_{l,m,n=0}^{\infty}e^{-\pi^2\left(\frac{l^2}{a^2}+\frac{m^2}{b^2}+\frac{n^2}{c^2}\right) \ndv{t} }c_{lmn}^2&\cos\left(\frac{\pi l x_1}{a}\right)\cos\left(\frac{\pi m y_1}{b}\right)\cos\left(\frac{\pi n z_1}{c}\right)\\
    &\cos\left(\frac{\pi l x_2}{a}\right)\cos\left(\frac{\pi m y_2}{b}\right)\cos\left(\frac{\pi n z_2}{c}\right)
\end{align*}
and the $c_{lmn}^2$ terms are the $L^2(\Omega)$-normalising constants chosen as follows:
\begin{align*}
    c_{lmn}^2=\begin{cases}
    \frac{8}{abc},  \ \text{if all $l,m,n$ are non-zero},\\
    \frac{4}{abc}, \ \text{if 2 of $l,m,n$ are non-zero}\\
    \frac{2}{abc}, \ \text{if 
exactly one of $l,m,n$ is non-zero},\\
 \frac{1}{abc}, \ \text{if $l,m,n$ are $0$.}
    \end{cases}
\end{align*}

\noindent
\textbf{Example: The Cylinder} \\
Here $\Omega=\{(x,y,z): x^2+y^2\le 1,0\le z\le 1\}$, although it is most convenient to use 
polar co-ordinates: 
$\Omega=\{(r,\phi,z): 0\le r\le 1, 0\le \phi\le 2\pi, 0\le z\le 1\}$.
Then the heat kernel is given by
\begin{align*}
&p(\ndv{t},(r_1,\phi_1,z_1),(r_2,\phi_2,z_2))=\sum_{m=0}^{\infty}\sum_{\sqrt{\lambda}\in (J_0')^{-1}(0)}c_{m0\lambda}^2e^{(-\lambda-\pi^2 m^2)\ndv{t}}J_0(\sqrt{\lambda}r_1)\cos(\pi m z_1)J_0(\sqrt{\lambda}r_2)\cos(\pi m z_2)\\
&+\sum_{m=0}^{\infty}\sum_{n=1}^{\infty}\sum_{\sqrt{\lambda}\in (J_n')^{-1}(0)\setminus\{0\}}c_{mn\lambda}^2e^{(-\lambda-\pi^2 m^2)\ndv{t}}J_n(\sqrt{\lambda}r_1)\cos(n\phi_1)\cos(\pi m z_1)J_n(\sqrt{\lambda}r_2)\cos(n\phi_2)\cos(\pi m z_2)
 \\
 &+\sum_{m=0}^{\infty}\sum_{n=1}^{\infty}\sum_{\sqrt{\lambda}\in (J_n')^{-1}(0)\setminus\{0\}}c_{mn\lambda}^2e^{(-\lambda-\pi^2 m^2)\ndv{t}}J_n(\sqrt{\lambda}r_1)\sin(n\phi_1)\cos(\pi m z_1)J_n(\sqrt{\lambda}r_2)\sin(n\phi_2)\cos(\pi m z_2).
 \end{align*}
 Here, $J_n$ the $n$th Bessel function of the first kind, 
and these constants need to be chosen to be 
\begin{align*}
c_{mn\lambda}^2=\frac{1}{(\frac{1+\delta^m_0}{2})(\pi+\delta^n_0\pi)\int_{0}^{1}J_n(\sqrt{\lambda}r)^2rdr}
\end{align*}

\noindent
\textbf{Example: The Sphere} \\
We set $\Omega=\{(x,y,z):x^2+y^2+z^2\le 1\}$, or in spherical co-ordinates, 
$\Omega=\{(r,\phi,\theta):0\le r\le 1, 0\le \phi\le 2\pi, 0\le \theta\le \pi\}$. Then the heat kernel is given by 
\begin{align*}
    p(\ndv{t},(r_1,\phi_1,\theta_1),(r_2,\phi_2,\theta_2))\\
    =\sum_{l=0}^{\infty}\sum_{m=0}^{\infty}\sum_{\sqrt{\lambda}\in (j'_l)^{-1}(0)}&e^{-\lambda \ndv{t}}c_{lm\lambda}^2j_l(\sqrt{\lambda} r_1)P^m_l(\cos(\theta_1))\sin(m\phi_1)j_l(\sqrt{\lambda} r_2)P^m_l(\cos(\theta_2))\sin(m\phi_2)\\
    &+e^{-\lambda \ndv{t}}c_{lm\lambda}^2j_l(\sqrt{\lambda} r_1)P^m_l(\cos(\theta_1))\cos(m\phi_1)j_l(\sqrt{\lambda} r_2)P^m_l(\cos(\theta_2))\cos(m\phi_2),
\end{align*}
where $j_l$ is the $l$th spherical Bessel function, and $P^m_l$ is an associated Legendre polynomial. The expression for $c_{lm\lambda}$ is quite complicated. However, our application of this heat kernel will involve imposing symmetry in $\phi$, in which case, it suffices to know what these constants are for $m=0$:
\begin{align*}
    c_{l0\lambda}^2=\frac{2l+1}{4\pi\int_{0}^{1}j_l(\sqrt{\lambda}r)^2r^2dr}
\end{align*}

\section{Application of the Sphere Heat Kernel}\label{sec:Applications}

\subsection{Evaluation of the Impulse Response Function}

\begin{figure}
 \centering
 \includegraphics[width=89mm]{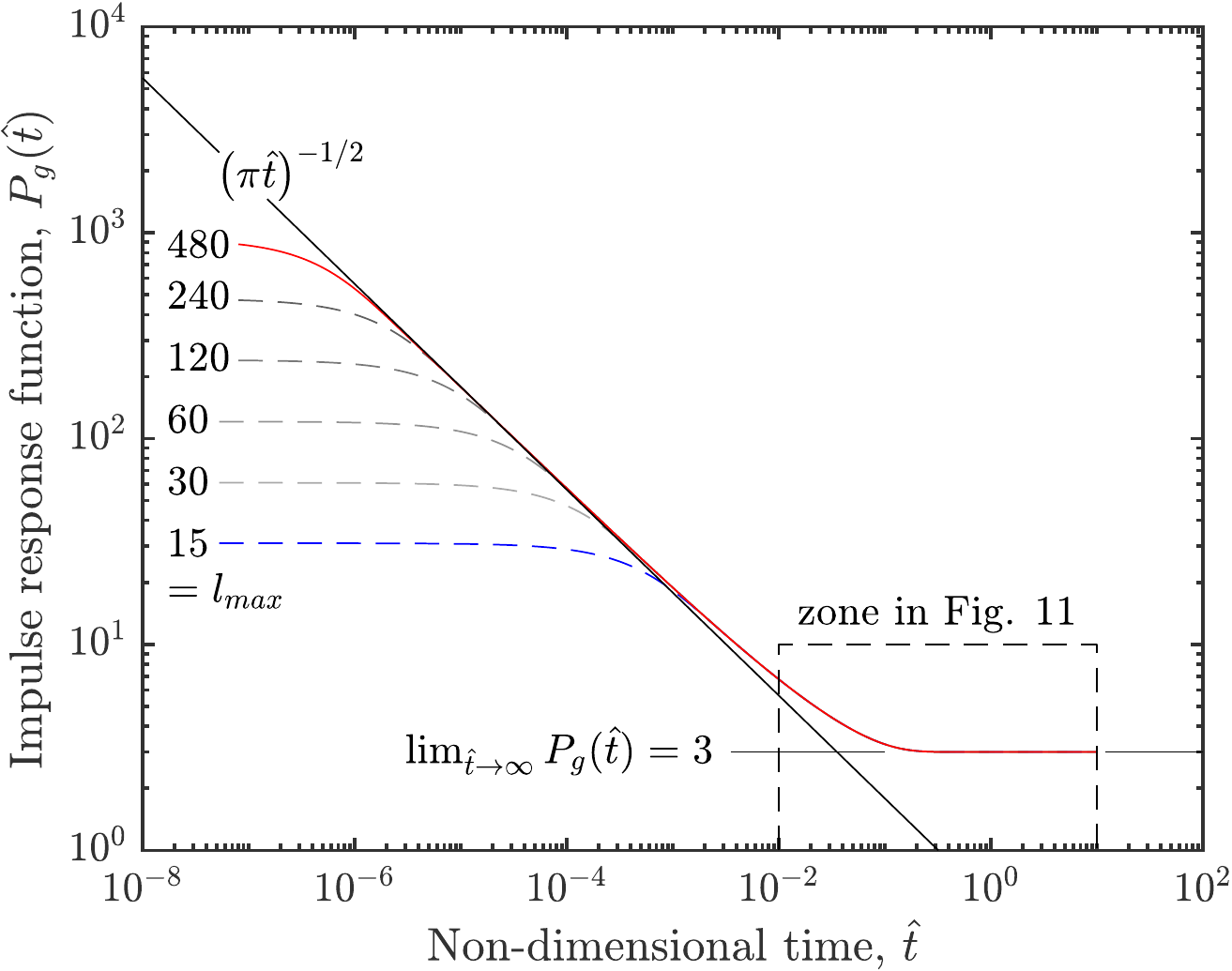}
 \caption{Sphere non-dimensional impulse response function $P_g(\ndv{t})$ with $g(\theta) = 1$ for several cases with the infinite sum terminated at $l_{max}$.}
 \label{fig:convergence}
\end{figure}

The non-dimensional impulse response function $P_g(\ndv{t})$ for the sphere with symmetry about the polar axis (symmetry in $\phi$) and with $g(\theta) = 1$ was numerically evaluated and results are presented in Fig.~\ref{fig:convergence}. 
For each $l$, the critical points of the $l$th spherical Bessel function $j_l$ were found using a single-variable nonlinear zero finding function (Matlab's \texttt{fzero} function) with the search range between the roots of the spherical Bessel function itself.
The normalising constants $c_{l 0 \lambda}^2$ were found by numerically evaluating the specified integral of the spherical Bessel function (using Matlab's \texttt{integral} function).

To achieve the results in Fig.~\ref{fig:convergence}, the infinite sum was terminated at several different maximum values of $l$ from $l_{max} = 15$ to $l_{max} = 480$, generally giving different values for $P_{g}(\ndv{t},l_{max})$ as indicated in Fig.~\ref{fig:convergence}.  In each case, the number of critical points of the spherical Bessel function included in the summation was also taken as the value of $l_{max}$.  Each plot for the different $l_{max}$ values consists of 200 logarithmically-space points in time.  At each of these values of time, the heat kernel $p(\ndv{t},(r_1,\theta_1),(r_2,\theta_2))$ with $r_1 = 1$ and $\theta_1 = 0$ was evaluated at 2001 uniformly distributed points around the boundary ($r_2 = 1$) from $\theta_2 = 0$ to $\theta_2 = \pi$. To evaluate the approximation of the impulse response function $P_g(\ndv{t},l_{max})$, spatial integration of the heat kernel on the boundary with $g(\theta) = 1$ was achieved using piece-wise polynomial (cubic spline) segments fitted to the data at the 2001 points.

\begin{figure}
 \centering
 \includegraphics[width=89mm]{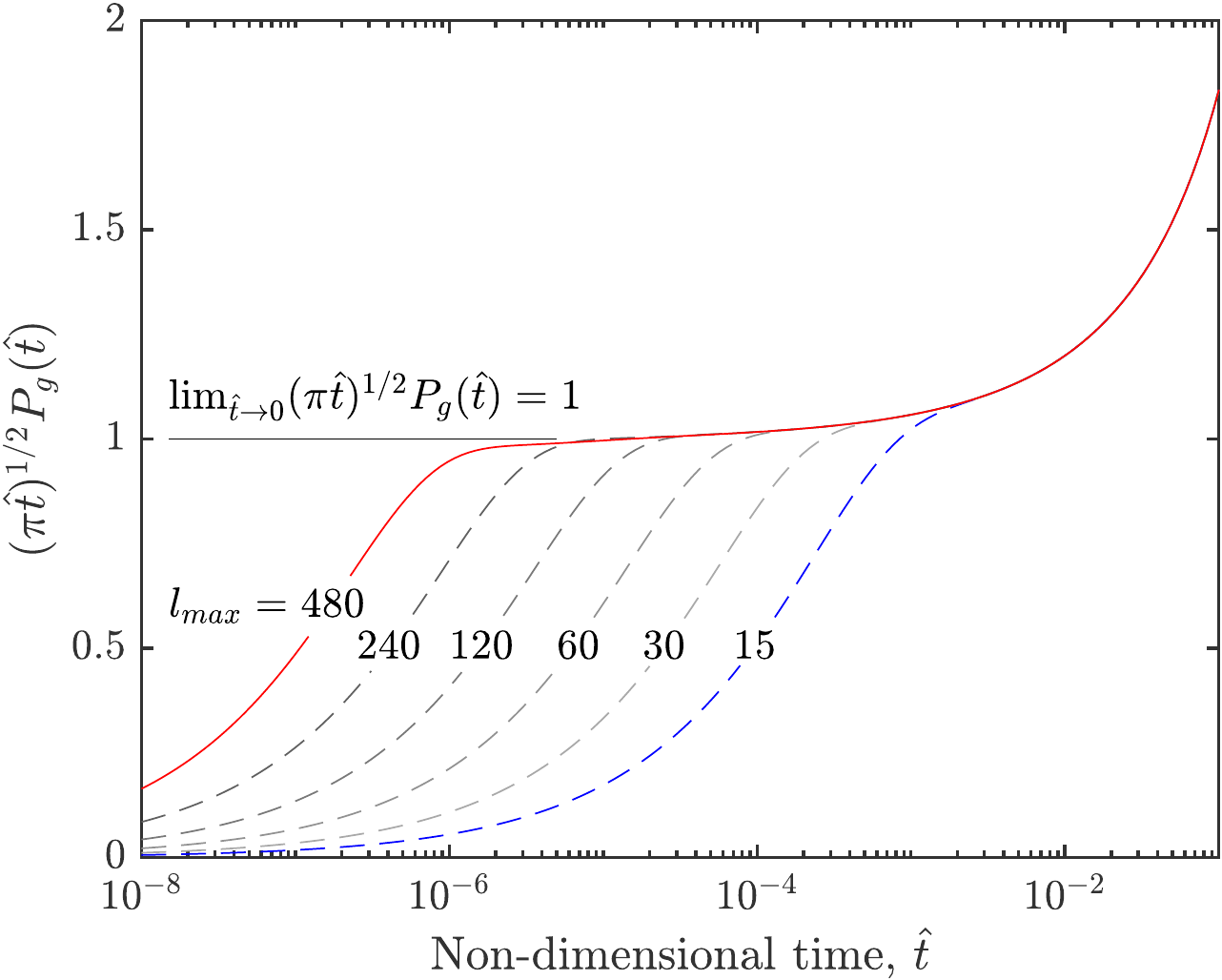}
 \caption{Sphere non-dimensional impulse response function normalised using the semi-infinite flat surface result giving, $(\pi \ndv{t})^{1/2} P_g(\ndv{t})$, with $g(\theta) = 1$ for several cases with the infinite sum terminated at $l_{max}$.}
 \label{fig:convergence_view2}
\end{figure}

Figure~\ref{fig:convergence_view2} presents a normalised view of the impulse response function and demonstrates that for sufficiently small values of $\ndv{t}$, the evaluated value of $P_g(\ndv{t},l_{max})$ falls below the value for the semi-infinite flat surface result $P_{g,si,fs}(\ndv{t}) = (\pi \ndv{t})^{-1/2}$.  However, with inclusion of additional terms in the summation (increasing the value of $l_{max}$), the evaluated magnitude of the impulse response function increases, and more closely approximates the value of $(\pi \ndv{t})^{-1/2}$.  Such convergence effects are also described in Table~\ref{tab:convergence} where the non-dimensional times $\ndv{t}$ are identified for several values of the ratio:
\begin{equation}
\nonumber
    \frac{P_{g}(\ndv{t},l_{max})}{P_{g}(\ndv{t},480)} \in \{ 0.995, 0.990, 0.980 \} .
\end{equation}
Values in Table~\ref{tab:convergence} demonstrate that at times $\ndv{t} \sim 10 \times 10^{-5}$, for heat kernel computations to produce results that remain within $\sim 1\,\%$ of the true impulse response function, about 60 terms need to be included in the summation.  Results in Table~\ref{tab:convergence} are also presented for non-dimensional times $\ndv{t}$ such that
\begin{equation}
\nonumber
    \frac{P_{g}(\ndv{t},l_{max})}{P_{g,si,fs}(\ndv{t})} \in \{ 1.005, 1.010, 1.020 \} ,
\end{equation}
which provides a measure for convergence of the computed impulse response function to the semi-infinite flat surface result.  With a total of 60 terms included in the summation, results are larger than the semi-infinite result by $\sim 1\,\%$ for times $\ndv{t} \sim 10 \times 10^{-5}$.

Examination of the computed results in Fig.~\ref{fig:convergence} and Fig.~\ref{fig:convergence_view2} suggests that two limits for the actual impulse response function exist.
\begin{enumerate}
    \item For short times: $\lim_{\ndv{t} \rightarrow 0 } (\pi \ndv{t})^{1/2}P_g (\ndv{t})  = 1$.
    \item For long times: $\lim_{\ndv{t} \rightarrow \infty } P_g (\ndv{t})  = 3$.
\end{enumerate} The limiting value for long times arises because of the ratio of the surface area to volume of the sphere with unit radius and it poses no computational issue as only a modest number of terms is required in the summation to achieve the result with high precision.  However, the short time limit represents some challenge.  While at any given $\ndv{t}$, increasing the value of $l_{max}$ tends to improve the quality of the approximation, computational precision limitations reduce the merit of including additional terms beyond $l_{max} \sim 120$. 

\begin{table}[h!]
  \begin{center}
    \caption{Values of $\ndv{t}$ for which certain $P_g$ ratios are $\sim 1.0$. }
    \label{tab:convergence}
    \begin{tabular}{r|ccc} 
          \hline
      $P_{g}(\ndv{t},l_{max})/P_{g}(\ndv{t},480)$ & $0.995$ & $0.990$ & $0.980$ \\
       $l_{max} =240$ & $0.54\times10^{-5}$ & $0.48\times10^{-5}$ & $0.42\times10^{-5}$ \\ 
      120 & $2.49\times10^{-5}$ & $2.16\times10^{-5}$ & $1.78\times10^{-5}$ \\ 
      60 & $10.77\times10^{-5}$ & $9.15\times10^{-5}$ & $7.49\times10^{-5}$ \\ 
      30 & $43.90\times10^{-5}$ & $36.83\times10^{-5}$ & $30.04\times10^{-5}$ \\
      15 & $174.18\times10^{-5}$ & $146.75\times10^{-5}$ & $119.57\times10^{-5}$ \\
    \hline
          $P_{g}(\ndv{t},l_{max})/P_{g,si,fs}(\ndv{t})$ & $1.005$ & $1.010$ & $1.020$ \\
           $l_{max} = 480$ & $2.57\times10^{-5}$ & $4.68\times10^{-5}$ & $13.39\times10^{-5}$ \\
           $240$ & $2.22\times10^{-5}$ & $4.59\times10^{-5}$ & $13.39\times10^{-5}$ \\
           $120$ & $3.00\times10^{-5}$ & $4.26\times10^{-5}$ & $13.26\times10^{-5}$ \\
           $60$ & $9.00\times10^{-5}$ & $10.06\times10^{-5}$ & $14.50\times10^{-5}$ \\
           $30$ & $28.28\times10^{-5}$ & $29.91\times10^{-5}$ & $34.47\times10^{-5}$ \\
           $15$ & $87.31\times10^{-5}$ & $90.58\times10^{-5}$ & $97.78\times10^{-5}$ \\
         \hline
    \end{tabular}
  \end{center}
\end{table}

\begin{figure}
 \centering
 \includegraphics[width=89mm]{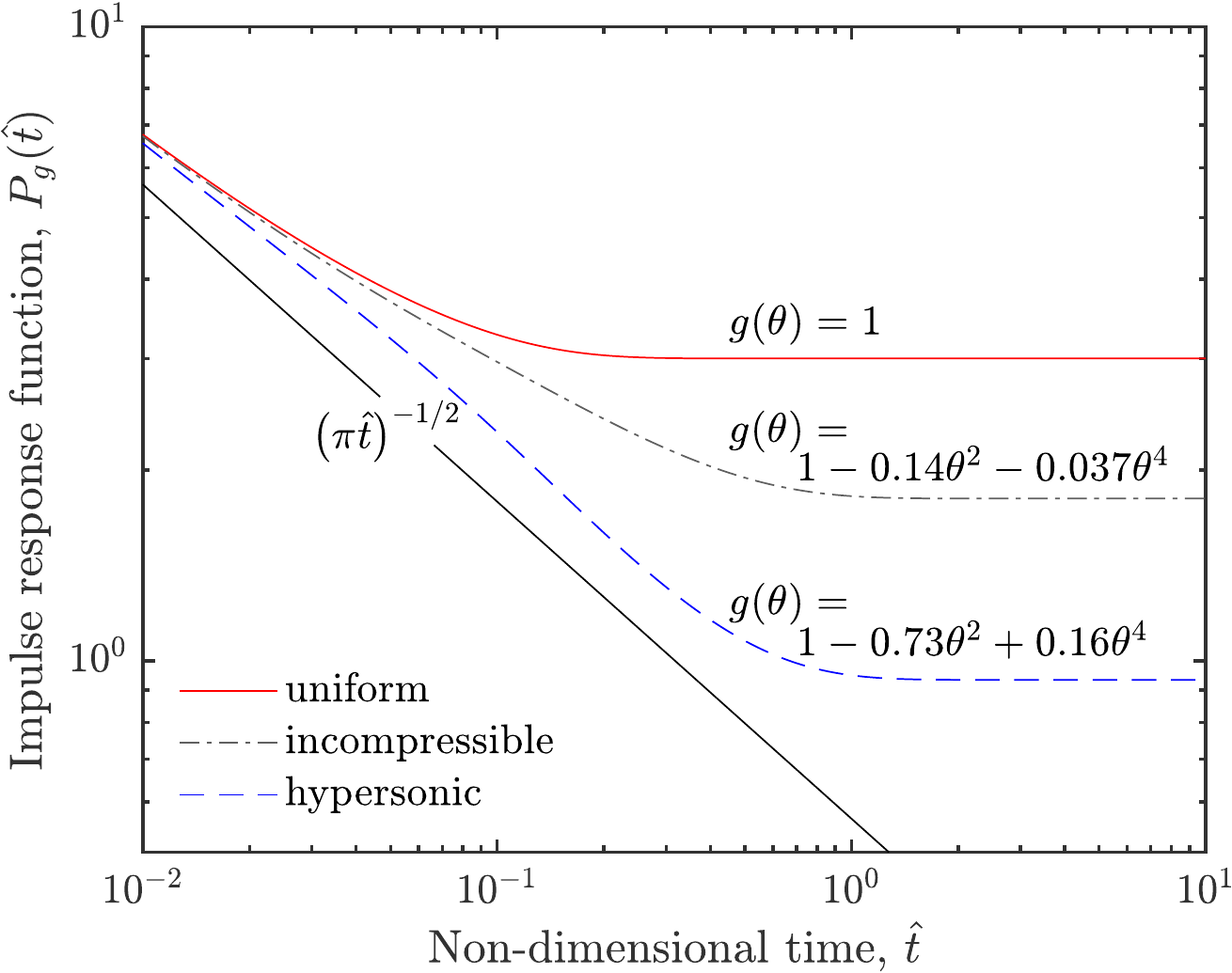}
 \caption{Sphere non-dimensional impulse response function $P_g(\ndv{t})$ for several different spatial variations of flux $g(\theta)$.}
 \label{fig:g_distribution}
\end{figure}

Impulse response function results for two cases in which the spatial variation of flux $g(\theta) \ne 1$ are shown in Fig.~\ref{fig:g_distribution}. These results were obtained in a similar manner to that described for the $g(\theta) = 1$ case with $l_{max} = 480$, which is also shown in Fig.~\ref{fig:g_distribution}.  Note that the range of non-dimensional times illustrated in Fig.~\ref{fig:g_distribution} is somewhat smaller than shown in Fig.~\ref{fig:convergence}.  The two spatial variation functions chosen for production of the results in Fig.~\ref{fig:g_distribution} correspond to the curve-fits used for the experimental data illustrated in Fig.~\ref{fig:sphere_data}: one for the incompressible flow case, and the other for the hypersonic flow case.  The hypersonic flow case with $g(\theta) = 1 - 0.73 \theta^2 + 0.16 \theta^4$ results in smaller values for the impulse response function at all times relative to the incompressible flow case with  $g(\theta) = 1 - 0.14 \theta^2 - 0.037 \theta^4$, and the magnitude of both of these functions is also smaller than that for the uniform case $g(\theta) = 1$ at all times, reflecting the relative magnitude of the respective distributions shown in Fig.~\ref{fig:sphere_data}.

\subsection{Impulse Response Filtering}

The method used to determine the heat flux from the measured surface temperature data involves an application of the Oldfield \cite{Oldfield2008} discrete-time impulse response filter technique.  With this method, the heat flux gauge is effectively treated as a linear, time-invariant system that can be characterised as using a discrete-time transfer function. To compute the discrete-time transfer function for the system, a pair of suitable, non-zero basis functions -- the system output corresponding to a particular system input -- must be defined.  Oldfield \cite{Oldfield2008} identifies suitable basis functions for several types of transient heat flux gauge using a step of heat flux at the gauge surface and the corresponding surface temperature history derived from one-dimensional heat conduction models for the gauge.  In the present work, we can apply the same methods, even though our heat conduction model is multi-dimensional.  

We have already illustrated what the impulse response function looks like for several different spatial variations of flux, Fig.~\ref{fig:g_distribution}. To obtain the surface temperature history corresponding to a heat flux step input at the measurement point, the impulse response function must be integrated with respect to time.  However, commencing the integration from $\ndv{t} = 0$ poses a difficulty because it is not possible to compute a reliable value for the heat kernel impulse response function as $\ndv{t} \rightarrow 0$.  Instead, it is recognised that $\lim_{\ndv{t} \rightarrow 0 } (\pi \ndv{t})^{1/2}P_g (\ndv{t})  = 1$, so the integration proceeds via the following steps. 

\begin{enumerate}
    \item Identify the number of terms to be included in the summation $l_{max}$ and specify an array of times $\ndv{t}_i$ at which the heat kernel impulse response function is evaluated, $P_g(\ndv{t}_i)$.
    \item Assess the computed impulse response values relative to the semi-infinite flat surface result and find the minimum value of time $\ndv{t}_i = \ndv{t}_n$ for which $P_g(\ndv{t}_i) > (\pi \ndv{t}_i)^{-1/2}$.
    \item Assign $P_g(\ndv{t}_{n-1}) = (\pi \ndv{t}_{n-1})^{-1/2}$ and then numerically integrate $P_g(\ndv{t}_i)$, but commence only from $t_{n-1}$.  In the present work this integration was achieved by fitting piece-wise polynomial segments (cubic splines) to the data and evaluating the integrated polynomials.
    \item The value of the constant that must be added to the approximation of the step function (the numerically-integrated $P_g(\ndv{t}_i)$ from Step 4) for $\ndv{t}_i > \ndv{t}_{n-1}$ is obtained from the analytical value for the semi-infinite flat surface: $2(\ndv{t}_{n-1}/\pi)^{1/2}$.
    \item If values of the step response function for $\ndv{t}_i \le \ndv{t}_{n-1}$ are required, these are also obtained from the semi-infinite flat surface result: $2(\ndv{t}_{t}/\pi)^{1/2}$.
\end{enumerate}

\begin{figure}
 \centering
 \includegraphics[width=89mm]{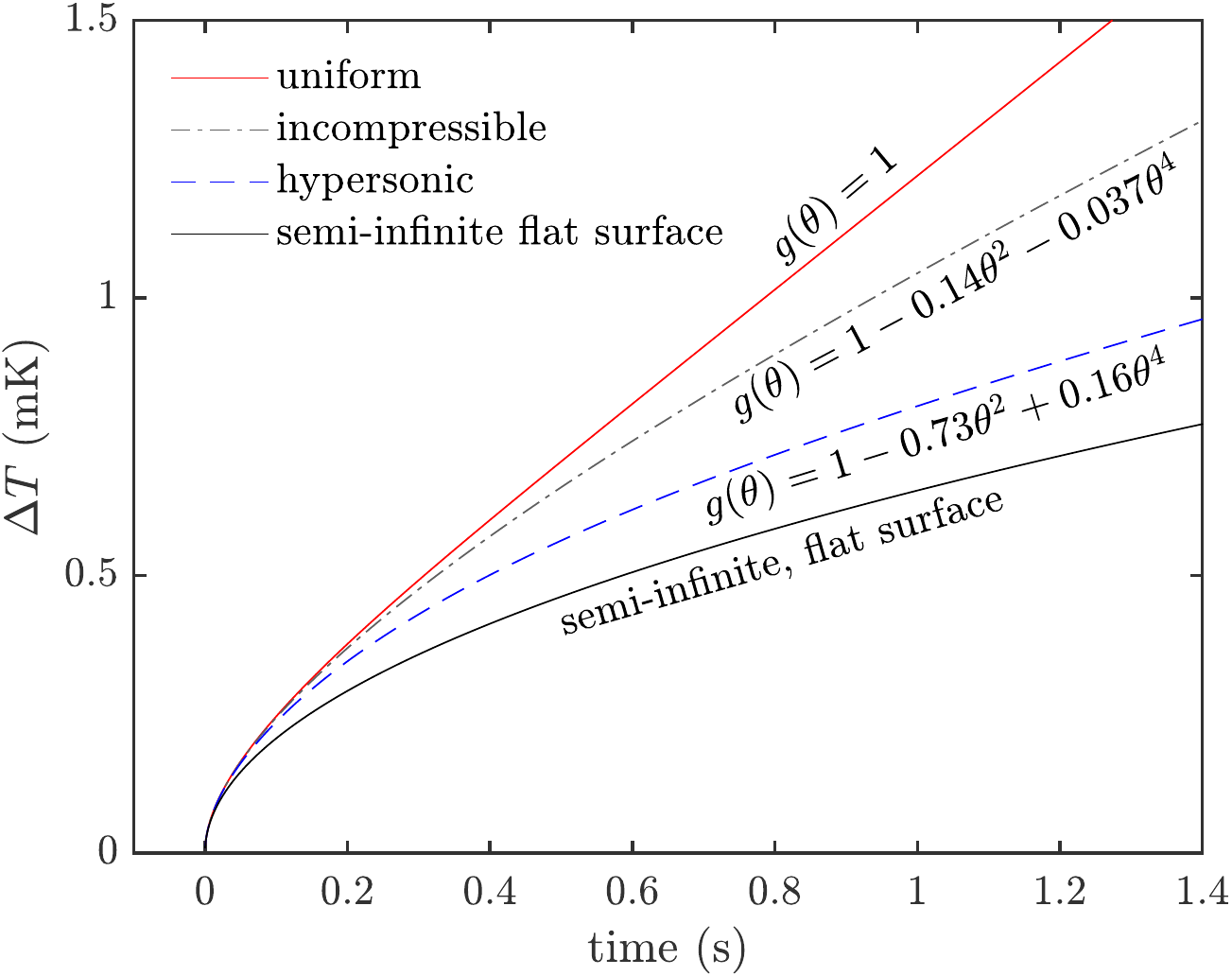}
 \caption{Temperature rise at the measurement point associated with a step input of heat flux, $q = 1$\,W/m$^2$ from $t = 0$ for several spatial variations of flux $g(\theta)$ and the semi-infinite flat surface case for the hot probe.}
 \label{fig:T_basis_functions}
\end{figure}

For the case of the hot probe, results from the integration of the non-dimensional impulse response are illustrated in Fig.~\ref{fig:T_basis_functions}, after scaling to give dimensional quantities.  The results for the cold probe are not shown because they differ from those of the hot probe by only a small amount; the small difference arises due to the temperature-dependent nature of the thermal properties.  For each probe, the substrate thermal properties are treated as constant and equal to the initial values at the flow onset.  The line labelled as the semi-infinite flat surface result is the well-recognised result given by
\begin{equation}
    \Delta T_{si,fs} = \frac{2 q_{step}}{\sqrt{\pi}\sqrt{\rho c k}} \sqrt{t}
\end{equation}
where $q_{step} = 1$\,W/m$^2$ is the value of the step input of heat flux that occurs at time $t = 0$.

Figure~\ref{fig:T_basis_functions} illustrates a larger temperature rise at the measurement location for the sphere with uniform heat flux around its surface than for the case of the semi-infinite flat surface, which is the expected result.  The incompressible and the hypersonic results, for which $g(\theta) \ne 1$, fall between the temperature rise for the uniform result and the semi-infinite flat surface result.  However, such results do not apply generally for $g(\theta) \ne 1$ since it is the form of the spatial distribution of the flux that plays a significant role.  In the two non-uniform cases considered here -- the incompressible and the hypersonic cases -- the peak heat flux occurs at the measurement point and the spatial variation of surface heat flux is gradual.

\subsection{Deduction of Heat Flux}

\begin{figure}
 \centering
 \includegraphics[width=89mm]{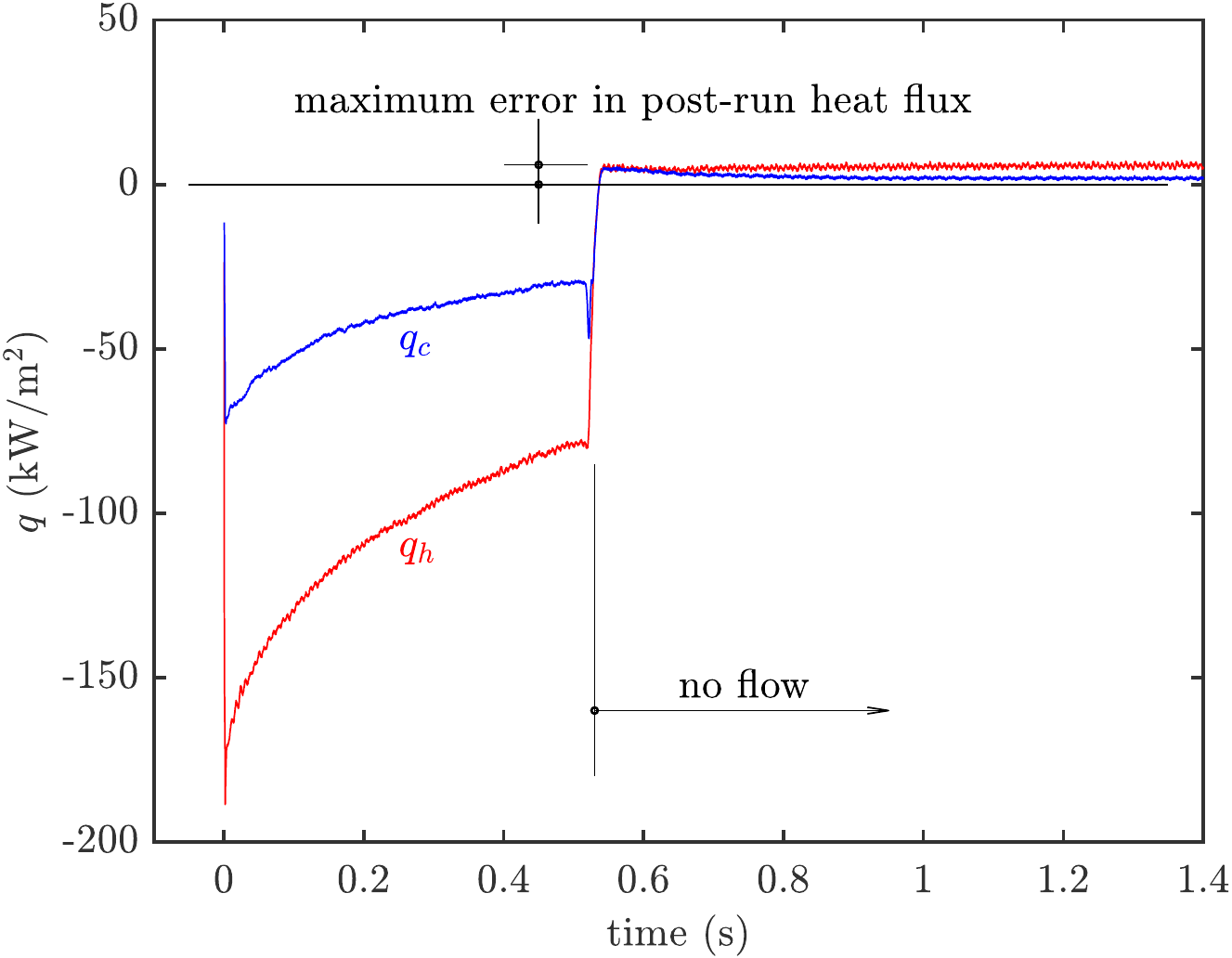}
 \caption{Heat flux results for the hot probe and the cold probe using the heat kernel approach.}
 \label{fig:heatflux_exp_hk_method}
\end{figure}

In the experiments, the flow around the hemispherical-nosed probes was subsonic, at a Mach number of approximately 0.4, so it is expected that the incompressible spatial variation function will represent the distribution of heat flux more accurately than the hypersonic variation.  Following the Oldfield method, discrete impulse response filters were established using, as basis functions, $q_{step} = 1$\,W/m$^2$ and the surface temperature history labelled \emph{incompressible} in Fig.~\ref{fig:T_basis_functions} for the hot probe, and a similar result for the cold probe.  These impulse response filters were then applied to the surface temperature histories for the hot and cold probes (as illustrated in Fig.~\ref{fig:p_and_T_exp}), yielding the results illustrated in Fig.~\ref{fig:heatflux_exp_hk_method}.

A comparison of Fig.~\ref{fig:heatflux_exp_2} and Fig.~\ref{fig:heatflux_exp_hk_method} demonstrates a substantial reduction in the heat flux error in the post-flow period achieved using the heat kernel approach.  In the case of the heat kernel method applied to the hot probe, the apparent heat flux over the post-flow period was $\sim 5.2$\,kW/m$^2$, and the corresponding value for the cold probe was $\sim3.3$\,kW/m$^2$.  In the case of the previous approach using corrections to the semi-infinite flat surface method, for the hot probe the value was $\sim13.3$\,kW/m$^2$ and for the cold probe the value was $\sim6.5$\,kW/m$^2$.  Therefore, errors are greatly reduced through the application of the heat kernel method, and a factor of 2 improvement is representative. 

The magnitude of the heat flux immediately prior to flow termination was $\sim 79.0$\,kW/m$^2$ in the case of the hot probe, and $\sim 29.5$\,kW/m$^2$ in the case of the cold probe.  The relative errors that exist in the deduced heat flux values at these times are estimated based on the apparent post-flow heat flux to be $\sim 7$\,\% in the case of the hot probe, and $\sim 11$\,\% in the case of the cold probe.  While the assumption of constant thermal properties for the probes will make a contribution to these errors, such a deficiency offers no explanation for the larger relative error in the case of the cold probe, because variable thermal property effects are related to temperature changes and these are smaller in the case of the cold probe relative to the hot probe.

Another possible contribution to the apparent error in the post-flow heat flux is the existence of a spatial variation of flux that is also a function of time. Although the flow conditions of experiments considered herein were essentially constant and were applied and removed in a step-like manner, some temporal dependence will arise because the local surface heat flux is driven by the temperature difference between the flow and the surface, scaled by the heat transfer coefficient.  In future investigations, it should be possible to extend the methods described herein to cases where the heat transfer coefficient, rather than the heat flux itself, has a prescribed spatial variation.

\section{Conclusion}

A new approach is introduced for analysis of transient heat transfer experiments in cases where the surface temperature history is measured at a single location and the relative spatial variation of heat flux in the vicinity of the measurement location is known.  
The new analysis is particularly valuable for configurations where the existence of multi-dimensional heat conduction effects within the substrate render the one-dimensional transient heat conduction approach invalid.
Semi-empirical approximations for correction of analyses built on the one-dimensional heat conduction assumption are also avoided in the new method.

Our approach demonstrates that an impulse response function suitable for characterisation of the measurement system can be obtained from the Neumann heat kernel.
Conceptually, development of the impulse response function involves the instantaneous deposition of a unit of energy at the measurement location on the boundary of the substrate, followed by identification of the temperature evolution around the boundary in the case of zero Neumann conditions.  
The impulse response function itself is then obtained by integrating the product of the relative spatial variation of heat flux and the surface temperature evolution around the boundary.

To demonstrate the practical usefulness of the techniques, an example involving hemispherical-nosed temperature probes that were operated in a subsonic flow is considered.  
Using previous experimental data for the relative spatial variation of heat flux around the boundary, and the temperature evolution from the Neumann heat kernel for the solid sphere, the impulse response function was computed. 
Transient heat flux results were then deduced from the measured surface temperature histories via a discrete impulse response filtering technique.
Errors arising in the new technique are demonstrated to be about a factor of two smaller than for an approach involving the correction of one-dimensional heat conduction results. 

Future applications for the technique are expected to arise in extended-duration transient experiments, where either the finite dimensions of the substrate or the gradients of heat flux at the boundary, or both, are significant.  
Future activity in this area could be directed towards extending the present method to the analysis of cases where the heat transfer coefficient, rather than the heat flux, has a specified spatial variation.

\bibliography{LateralConduction}

\end{document}